\documentclass[11pt, draftclsnofoot, onecolumn]{IEEEtran}

\usepackage{pifont}
\usepackage{cite}
\usepackage[dvips]{epsfig}
\usepackage{graphicx}
\usepackage{calligra}
\usepackage[T1]{fontenc}
\usepackage{bbold}
\usepackage{mathrsfs}

\usepackage{subfigure}
\usepackage{color}
\usepackage{amsmath}
\usepackage{cases}

\usepackage{amssymb}
\usepackage[justification=centering]{caption}
\usepackage{framed}

\usepackage{subfigure}
\usepackage{color}
\usepackage{amsmath}
\usepackage{bm}
\usepackage{amssymb}
\usepackage{amsbsy}
\usepackage{bbm}
\usepackage{dsfont}
\usepackage[normalem]{ulem}
\usepackage{xcolor,cancel}
\usepackage{enumerate}
\usepackage{paralist}

\usepackage{algorithm, algorithmic}

\usepackage{amsthm}

\newtheorem{theorem}{Theorem}

\newtheorem{condition}{Condition}

\newtheorem{corollary}{Corollary}

\newtheorem{definition}{Definition}

\newtheorem{proposition}{Proposition}

\newtheorem{lemma}{Lemma}

\newtheorem{remark}{Remark}

\newtheorem{claim}{Claim}

\newtheorem{assumption}{Assumption}

\DeclareMathOperator{\esssup}{ess\,sup}

\newcommand{\bW}{\bf{W}}

\newcommand{\bbE}{\mathbbm{E}}

\newcommand{\bbP}{\mathbbm{P}}

\newcommand{\cM}{\mathcal{M}}

\newcommand{\cG}{\mathcal{G}}

\newcommand{\cF}{\mathcal{F}}
\newcommand{\cE}{\mathcal{E}}

\newcommand{\cA}{\mathcal{A}}
\newcommand{\cN}{\mathcal{N}}

\newcommand{\cO}{\mathcal{O}}

\newcommand{\cS}{\mathcal{S}}

\newcommand{\bone}{\mathbbm{1}}

\newcommand{\bxi}{{\boldsymbol{\xi}}}

\newcommand{\bgamma}{{\boldsymbol{\gamma}}}
\newcommand{\cbgamma}{{\check{\boldsymbol{\gamma}}}}
\newcommand{\cgamma}{{\check{{\gamma}}}}
\newcommand{\tbgamma}{{\widetilde{\boldsymbol{\gamma}}}}
\newcommand{\tgamma}{{\widetilde{{\gamma}}}}

\newcommand{\bGamma}{{\boldsymbol{\Gamma}}}
\newcommand{\cbGamma}{{\check{\boldsymbol{\Gamma}}}}

\newcommand{\tbGamma}{{\widetilde{\boldsymbol{\Gamma}}}}

\newcommand{\cu}{{\check{u}}}

\newcommand{\cw}{{\check{w}}}

\newcommand{\tw}{{\widetilde{w}}}

\newcommand{\cW}{{\check{W}}}

\newcommand{\tW}{{\widetilde{W}}}

\newcommand{\ignore}[1]{{}}

\newcounter{parentalgorithm}

\ifCLASSINFOpdf
\else
\fi

\begin{document}
%
\title{Consensus-based Distributed Quickest Detection of Attacks with Unknown Parameters}
%
%
%

\author{{Jiangfan Zhang,~\IEEEmembership{Member,~IEEE},  Xiaodong Wang,~\IEEEmembership{Fellow,~IEEE}
}
\thanks{J. Zhang and X. Wang are with the Department of Electrical Engineering, Columbia University, New York, NY 10027 USA (e-mail: jiangfan.zhang@columbia.edu; wangx@ee.columbia.edu).}

}


\maketitle

\begin{abstract}

	
Sequential attack detection in a distributed estimation system is considered, where each sensor successively produces one-bit quantized samples of a desired deterministic scalar parameter corrupted by additive noise. The unknown parameters in the pre-attack and post-attack models, namely the desired parameter   to be estimated   and the injected malicious data at the attacked sensors pose a significant challenge for designing a computationally efficient   scheme   for each sensor to detect the occurrence of attacks by only using local communication   with   neighboring sensors.  The generalized Cumulative Sum (GCUSUM) algorithm is considered, which replaces the unknown parameters with their maximum likelihood estimates in the CUSUM test statistic.   For the problem under consideration,   a sufficient condition is provided under which the expected false alarm period   of the GCUSUM   can be guaranteed to be larger than any given value. Next, we consider the distributed implementation of the GCUSUM. We first propose an alternative test statistic which is asymptotically equivalent to that of GCUSUM. Then based on the proposed alternative  test statistic and running consensus algorithms, we propose a distributed approximate GCUSUM algorithm which   significantly reduce the prohibitively high  computational complexity of the centralized GCUSUM. Numerical results show that the distributed approximate GCUSUM algorithm can provide a performance that is comparable to the centralized  GCUSUM.  

\end{abstract}




\section{Introduction}
\label{Section_Introduction}

Applications of sensor networks for parameter estimation, ranging from inexpensive commercial systems to complex military and homeland defense surveillance systems, have been extensively studied in recent   literature   \cite{akyildiz2002survey}. Typically, large-scale sensor networks consist of low-cost and spatially distributed sensors with limited battery power and low computing   capability,   which makes the system vulnerable to cyberattacks by adversaries. This has led to great interest in studying the vulnerability of sensor networks in various applications, see \cite{li2005robust, cui2012coordinated, vempaty2013distributed, nadendla2014distributed, zhang2015Asymptotically, alnajjab2015attacks,  zhang2017functional, zhang2017attack} and the references therein.  

Recently, one-bit quantization has been increasingly gaining attention in the applications of sensor networks and other systems like massive MIMO communication systems, see \cite{papadopoulos2001sequential, xiao2006distributed2, ribeiro2006bandwidth1, vempaty2013localization, mollen2017uplink, li2017channel} for instance. The application of one-bit quantization in such systems is partially motivated by the fact that the power consumption of the analog-to-digital (ADC) converters grows exponentially with the number of quantization bits \cite{walden1999analog}. One-bit ADC consists of a simple comparator which does not require automatic gain control or a highly linear amplifier, and therefore can be implemented with very low cost and power consumption \cite{mo2015capacity}.   This paper   considers a spatially distributed sensor network estimation system employing one-bit quantization under the assumption that the measurements from a subset of sensors are under the threat of attacks. 


There are primarily two types of sensor network architectures, depending on whether or not there exists a fusion center (FC). In the presence of an FC, all sensor data is collected and processed at the FC. Thus, one of major challenges associated with this type of sensor networks arises from the communication burden from sensors to the FC, especially when sensors are widely deployed and far away from the FC. For the other type of sensor networks without an FC, which are referred to as distributed sensor networks, the sensors are connected to their neighboring sensors, which allow them to exchange data using local short-range communication. In a distributed sensor network, each sensor needs to make its own inference based on its own information and the information received from its neighboring sensors. The information aggregation in distributed sensor networks has been   extensively   studied. Most of the existing literature focuses on the fixed-sample-size paradigm, where every sensor starts with some samples and aims to obtain the average of all samples in the distributed sensor network through inter-sensor information exchange. The most popular information exchange protocols are the ``consensus algorithm'' and ``gossip algorithm'', whose comprehensive surveys can be found in \cite{olfati2007consensus} and \cite{dimakis2010gossip}, respectively. In these works, a new sample is not allowed to enter into the system during the process of inter-sensor information exchange, and hence they are only relevant to the fixed-sample-size inference problems.

In distributed sequential inference problems, with successively arriving samples, the more practical scenario is that the sampling and information aggregation processes take place simultaneously, or at least in comparable time-scales. Under this setup, \cite{kar2013consensus} proposed the ``consensus + innovation'' approach for distributed recursive parameter estimation; \cite{braca2008enforcing} investigated the problem of tracking a stochastic
process using a ``running consensus'' algorithm.   The distributed sequential change-point detection  based on running consensus algorithms has not been well investigated.  

In this paper, we consider the problem of detecting the occurrence of attacks in a distributed sensor network estimation system with quantized data as quickly as possible while the expected false alarm period is under control. The sequential change-point detection, also known as quickest detection, which minimizes the expected detection delay subject to certain performance constraint on the expected false alarm period, suits well to this kind of problems. Various  change-point detection methods have been developed based on the well-known Cumulative Sum (CUSUM) algorithm \cite{basseville1993detection, tartakovsky2014sequential}.   However, most of the existing approaches either focus on the case where no unknown parameter exists, or require an FC that collects information (raw data or statistics) from all sensors to implement the algorithm, since the test statistic of the CUSUM algorithm accumulates the log-likelihood ratio of each data nonlinearly.  Here, we propose a new distributed change-point detection procedure based on running consensus algorithms, which allows each sensor in the network to  detect the occurrence of  attacks even though there are unknown parameters.  

\subsection{Summary of Results}

In this work, we consider the sequential attack detection in  a distributed sensor network where each sensor successively makes one-bit quantized samples of a desired deterministic scalar parameter corrupted by
additive noise. The network is composed of two types of sensors. One type of sensors have a very high level of security and thereby are guaranteed to be tamper-proof. The rest of the sensors are insecure, which are subject to attacks. In practice, the secure sensors can be well protected, built with powerful chips, and thereby highly sophisticated encryption algorithms and security procedures can be implemented.
At some time instant, the adversaries may launch attacks at the insecure sensors simultaneously by injecting different malicious data   into these    sensors.  It is well-known that the CUSUM algorithm is minimax optimal if the pre-attack and post-attack statistical models of the observations can be specified \cite{moustakides1986optimal}.  However, neither pre-attack nor post-attack  models  can be specified in our considered problem, since there are unknown parameters in both statistical models, namely the desired  parameter   to be estimated   and the injected malicious data. In light of this, we consider a CUSUM-type algorithm based on the generalized likelihood ratio (GLR), called the generalized CUSUM (GCUSUM) algorithm, which replaces the unknown parameters with their maximum likelihood estimates in the CUSUM test statistic.

It is shown that unlike the case where the pre-attack statistical model of the observations can be specified, the secure data also contributes to the test statistic of the GCUSUM due to the uncertainty of the desired parameter in the pre-attack statistical model. In addition, unlike the test statistic of the CUSUM for the case where the pre-attack and post-attack statistical models of the observations can be specified, the test statistic of the GCUSUM cannot be expressed in a recursive form evolving over time, which hinders us from efficiently implementing the GCUSUM. 

  Moreover,   the expected false alarm period may not be controlled for the GCUSUM in general. In order to overcome this, before carrying out the inference task, we secure the system for a period of time during which the whole system is tamper-proof. A sufficient condition on the length of the secure time is provided under which the expected false alarm period   of the GCUSUM test   can be guaranteed to be larger than any prescribed value. Furthermore, we show that if the length of secure time or the number of sensors in the network are large enough, the false alarm period   the GCUSUM test   can be guaranteed to be arbitrarily large with   arbitrarily   high probability. 


  It is worth mentioning that there are two major difficulties in the implementation of the GCUSUM for the problem under consideration.  First, the computation of the test statistic of the GCUSUM requires centralized processing which brings about a significant communication overhead in large-scale spatially distributed sensor networks.  Second, even if centralized processing is allowed, the computational complexity of the test statistic of the GCUSUM is prohibitively high.
Motivated by these facts,   we consider the distributed computation of the test statistic of the GCUSUM. 
Due to the procedure for estimating the unknown parameters and the mixed-integer programming in the computation of the test statistic of the GCUSUM, it is impossible to implement the GCUSUM based on running consensus algorithms, since the running consensus algorithms essentially can only calculate the weighted sum of local statistics at the sensors. To facilitate the distributed implementation of the GCUSUM, we first propose an alternative test statistic as a substitute for the test statistic of the GCUSUM, which is shown to be asymptotically equivalent to that of the GCUSUM when the secure time is sufficiently long. Similar to the test statistic of the GCUSUM, this alternative test statistic can be written as a sum of  statistics which are contributed by the secure and insecure data, respectively. 

For the statistics contributed by the secure data, we propose a running consensus algorithm which enables each sensor to produce a local estimate of each statistic. We show that the estimation error can be bounded from above almost surely when the secure time is sufficiently long. Moreover, this upper bound on the estimation error decreases to   zero   as the number of message-passing   steps   in the proposed algorithm increases. To compute the statistic contributed by the insecure data in a distributed and computationally efficient way, we first relax the constraint   of   simultaneous attack and   assume   that the attack times at   different  sensors   can be   different.   Then   the statistic can be rewritten in a recursive form which can be calculated at each sensor by employing running consensus algorithms. 

\subsection{Related Work}

  Attack   detection in sensor networks using quantized data has been widely
investigated, see \cite{li2005robust, cui2012coordinated, vempaty2013distributed, nadendla2014distributed, zhang2015Asymptotically, alnajjab2015attacks, zhang2017functional, zhang2017attack} and the references therein.  Most of these works investigate the problem of detecting attacks    under   the fixed-sample-size paradigm, while this paper focuses on the sequential attack detection.    Quickest   detection in sensor networks has also gained a great deal of attention in recent literature, see \cite{veeravalli2001decentralized, mei2008asymptotic, tartakovsky2008asymptotically,  hadjiliadis2009one, raghavan2010quickest, fellouris2016second, huang2011defending, xie2013sequential,  li2015quickest,  liu2017distributed} for instance. However, most of these works require an FC to process all   local sensor   statistics centrally; while in this work, we consider the sequential attack detection in  a distributed sensor network where no FC exists and each sensor needs to make its own decision based on its own   data   and the received information from its neighboring sensors.

The closest works to this paper are \cite{braca2011consensus, stankovic2011distributed, ilic2012consensus, liu2017distributed}, which all focus on the consensus based change-point detection in distributed sensor networks. Still, there are major differences. In \cite{stankovic2011distributed}, \cite{braca2011consensus} and \cite{liu2017distributed}, the authors assume that the pre-change and post-change distributions are completely known; while in this work, none of them are completely specified. In \cite{ilic2012consensus}, a geometric moving average algorithm is employed for detecting the change; while in this work, we are primarily interested in  CUSUM-type algorithms.

A related but different problem, the distributed sequential hypothesis test based on consensus, has been studied in \cite{sahu2016distributed, li2016order},   where   the sequential probability ratio test   (SPRT) is employed    to resolve two hypotheses, which is different from the   change detection problem.  

The remainder of the paper is organized as follows. The
problem statement and background are described in Section \ref{Section_System_Model}. The
sequential attack detection based on GCUSUM is introduced  in Section \ref{Section_GCUSUM}. In Section \ref{Section_DCUSUM}, we propose a distributed approximate   GCUSUM   algorithm which can be implemented through running consensus algorithm. In Section \ref{Section_Numerical_Results},  numerical results are provided to illustrate the performance of our proposed approach. Finally, Section \ref{Section_Conclusions} provides our conclusions.

\section{Problem Statement and Background}
\label{Section_System_Model}

\subsection{Problem Statement}

We consider a  sensor network consisting of $N$ spatially distributed sensors, each sequentially making observations of a  deterministic scalar parameter $\theta$ corrupted by additive noise. At the $k$-th time instant, the signal presented to the $j$-th sensor is described by
\begin{equation} \label{x_j_k_model}
x_j^{\left( k \right)} = \theta  + n_j^{\left( k \right)},
\end{equation}
where $n_j ^{(k)}$ denotes the additive noise sample with common cumulative distribution function (cdf) $F(\cdot)$ and $\{n_j^{(k)}\}$ is an independent and identically distributed sequence. We assume that the cdf $F(\cdot)$ is invertible. 

Each sensor employs a common one-bit quantizer to covert the analog signal $x_j^{\left( k \right)} $ to the quantized observation $u_j^{\left( k \right)} $ in the   following way  
\begin{equation} \label{quantized_data}
u_j^{\left( k \right)}  = {\bone}\{x_j^{\left( k \right)}  \in (\tau, \infty)\},
\end{equation}
where $\tau$ is the threshold of the quantizer.  We assume that each sensor   can only    exchange information with its neighboring sensors, and attempts to  produce its estimate of the unknown parameter $\theta$. The inter-sensor communication links determine the topology of the sensor network, which can be represented by an undirected graph $\cG = \{ \cN,  \cE \}$  with $\cN$ being the set of sensors and $\cE$ the set of edges. 

We assume that there exist a subset $\cS$ of sensors which are secure, and we use $\cA = \cN \backslash \cS$ to denote the set of insecure sensors. The numbers of sensors in $\cS$ and $\cA$ are denoted by $N_{\cS}$ and $N_{\cA}$, respectively. It is worth mentioning that $\cS$ can be empty. Suppose that after a period of secure time $M$, some adversaries launch data injection attacks at some time $t_a>M$ at insecure sensors, hoping to cause the sensor network system to reach an inaccurate estimate of $\theta$. We consider the worst possible scenario regarding the number of corrupted sensors, and hence, our analysis focuses on the case where the set  of all insecure sensors are corrupted \cite{bayraktar2016efficient}. Accordingly, 
  for any $j \in \cA$, the observations become  
\begin{equation}
\left\{ \begin{array}{l}
x_j^{\left( k \right)} = \theta  + n_j^{\left( k \right)}, \qquad  \quad k < {t_a}\\
x_j^{\left( k \right)} = \theta  + {\mu _j} + n_j^{\left( k \right)}, \quad k \ge {t_a},
\end{array} \right.
\end{equation}
where $\mu_j$ is an unknown deterministic injected data at the $j$-th sensor which satisfies
\begin{equation} \label{mu_constraint}
\mu_j \ge b >0, \quad \forall j \in \cA.
\end{equation}
The quantity $b$ is a prescribed value, setting the lower bound for the measurement change that draws security attentions. As such, our goal here is to detect the attacks  as soon as possible after their occurrence at $t_a$. The quickest detection technique, that exploits the statistical difference before and after the
change-point, provides a suitable framework to achieve this goal. For the sake of notational simplicity, we use $\{\check{u}_j^{(m)}\}$ to denote the quantized data  in the period of secure time, that is, ${\check{u}}_j^{(m)} = {u}_j^{(m)}$ for all $j$ and $m=1,2,...,M$, and we relabel ${u}_j^{(M+k)}$ in (\ref{quantized_data}) as ${u}_j^{(k)}$  for all $j$ and $k\ge 1$.

\subsection{Quickest Detection}

In quickest detection, we observe samples sequentially in time, and then stop when a predesigned rule declares a change. Unlike the fixed-sample-size schemes that lay more attention on the detection power, the sequential change detector aims at minimizing the expected detection delay after the change-point. The commonly used performance measure is the worst-case expected detection delay, proposed by Lorden \cite{lorden1971procedures}
\begin{equation} \label{lorden_formulation}
J\left( T \right) = \mathop {\sup }\limits_t  {\esssup _{{{\cal F}_t}}}{\bbE_t}\left[ {{{\left( {T - t} \right)}^ + }\left| {{{\cal F}_t}} \right.} \right],
\end{equation}
where $T$ is a stopping time variable corresponding to a certain detection scheme and $\cF_t$ is the filtration generated by all the observations up to time $t$. 
The expectation ${\bbE_t}$ is evaluated with respect to the post-change probability measure conditioned on the change-point $t$ and the all  observations up to time $t$. The essential supremum is obtained over ${{\cal F}_t}$, yielding the least favorable situation for the expected detection delay. The supremum in (\ref{lorden_formulation}) is obtained over $t$, meaning that the change occurs at such a point that the expected detection delay is maximized. To summarize, $J(T)$ characterizes the expected detection delay for the worst possible history of observations before the change-point. While the small expected detection delay under attack results in timely alarmed reaction, the running length under unattacked data, on the other hand, needs to be guaranteed to avoid frequent false alarms. To this end, the sequential change detection problem is formulated as follows:
\begin{equation} \label{changepoint_detection_problem_formulation}
\mathop {\inf }  \limits_T   J\left( T \right)  {\text{  subject to  }}   {\bbE_\infty }\left\{ T \right\} \ge \kappa.
\end{equation}
Note that  the expectation ${\bbE_{\infty}}$ is evaluated with respect to the probability measure where no change occurs, i.e., $t=\infty$, and $\kappa$ is a prescribed constant which specifies the required lower bound on the expected false alarm period. To proceed with our attack
detection problem, we denote the probability measures under no attack and under attack as $\bbP_0$ and $\bbP_1$, respectively. If all the parameters $\theta$ and $\{\mu_j\}$ are known, the quickest detection problem in (\ref{changepoint_detection_problem_formulation}) is optimally solved by the well-known CUSUM test \cite{moustakides1986optimal}
\begin{equation} \label{cusum_test}
{T_C} = \min \left\{ {K:\mathop {\max }\limits_{1 \le k \le K} \sum\limits_{i = k}^K {\sum\limits_{j \in {\cal A}} {\log \frac{{{\bbP_1}\left( {u_j^{(i)}\left| {\theta,  {\mu _j}} \right.} \right)}}{{{\bbP_0}\left( {u_j^{(i)}\left| \theta  \right.} \right)}} \ge h} } } \right\},
\end{equation}
where the threshold $h$ is determined by the constraint in (\ref{changepoint_detection_problem_formulation}). For a given $K$, the value of $k$ which maximizes the test statistic in (\ref{cusum_test}) can be considered as the estimate of attack time. It is well-known that the nonlinear accumulation of the log-likelihood ratios in (\ref{cusum_test}) can be written in a recursive way, and hence can be easily implemented in practice with low complexity \cite{basseville1993detection, tartakovsky2014sequential}. 

Note that in our attack detection problem, the parameters $\theta$ and $\{\mu_j\}$ are unknown and need to be estimated, as estimating $\theta$ is the essential task of the network. To address this, this paper resort to the generalized likelihood   ratio (GLR) method by replacing   the unknown parameters with their maximum likelihood estimates (MLE) \cite{basseville1993detection, tartakovsky2014sequential}.

\section{Sequential Attack Detection Based on Generalized CUSUM Test}
\label{Section_GCUSUM}

\subsection{Generalized CUSUM Test}

Upon replacing the unknown parameters $\theta$ and $\{\mu_j\}$ with their MLEs, the generalized CUSUM test can be written as
\begin{equation} \label{GCUSUM_define}
{T_G} = \min \left\{ {K:\mathop {\max }\limits_{1 \le k \le K} \Lambda _G^{(k,K)} \ge h} \right\},
\end{equation}
where $T_G$ is the stopping time corresponding to the generalized CUSUM test, and the statistic $\Lambda _G^{(k,K)} $ is given by
\begin{align} \notag
\Lambda _G^{(k,K)} & \buildrel \Delta \over = \ln \frac{{\mathop {\sup }\limits_{\theta ,\{ {\mu _j} \ge b\} } \prod\limits_{m = 1}^M {\prod\limits_{j \in {\cal N}} {{\bbP_0}\left( {{\check{u}}_j^{(m)}\left| \theta  \right.} \right)} } \prod\limits_{i = 1}^K {\prod\limits_{j \in {\cal S}} {{\bbP_0}\left( {u_j^{(i)}\left| \theta  \right.} \right)} } \prod\limits_{i = 1}^{k - 1} {\prod\limits_{j \in {\cal A}} {{\bbP_0}\left( {u_j^{(i)}\left| \theta  \right.} \right)} } \prod\limits_{i = k}^K {\prod\limits_{j \in {\cal A}} {{\bbP_1}\left( {u_j^{(i)}\left| {\theta ,{\mu _j}} \right.} \right)} } }}{{\mathop {\sup }\limits_\theta  \prod\limits_{m = 1}^M {\prod\limits_{j \in {\cal N}} {{\bbP_0}\left( {{\check{u}}_j^{(m)}\left| \theta  \right.} \right)} } \prod\limits_{i = 1}^K {\prod\limits_{j \in {\cal N}} {{\bbP_0}\left( {u_j^{(i)}\left| \theta  \right.} \right)} } }}\\ \label{GCUSUM_statistic}
&  = \mathop {\sup }\limits_{\theta ,\{ {\mu _j} \ge b\} } {f_1}\left( {\theta ,\{ {\mu _j}\} } \right) - \mathop {\sup }\limits_\theta  {f_0}\left( {\theta } \right)
\end{align}
  with  
\begin{align} \notag
{f_1}\left( {\theta ,\{ {\mu _j}\} } \right) & \buildrel \Delta \over = \sum\limits_{m = 1}^M {\sum\limits_{j \in {\cal N}} {\ln {\bbP_0}\left( {{\check{u}}_j^{(m)}\left| \theta  \right.} \right)} }  + \sum\limits_{i = 1}^K {\sum\limits_{j \in {\cal S}} {\ln {\bbP_0}\left( {u_j^{(i)}\left| \theta  \right.} \right)} } \\ \label{f1_define}
& \quad + \sum\limits_{i = 1}^{k - 1} {\sum\limits_{j \in {\cal A}} {\ln {\bbP_0}\left( {u_j^{(i)}\left| \theta  \right.} \right)} }  + \sum\limits_{i = k}^K {\sum\limits_{j \in {\cal A}} {\ln {\bbP_1}\left( {u_j^{(i)}\left| {\theta ,{\mu _j}} \right.} \right)} },
\end{align}
\begin{equation} \label{f0_define}
{f_0}\left( \theta  \right) \buildrel \Delta \over = \sum\limits_{m = 1}^M {\sum\limits_{j \in {\cal N}} {\ln {\bbP_0}\left( {{\check{u}}_j^{(m)}\left| \theta  \right.} \right)} }  + \sum\limits_{i = 1}^K {\sum\limits_{j \in {\cal N}} {\ln {\bbP_0}\left( {u_j^{(i)}\left| \theta  \right.} \right)} }.
\end{equation}


By defining
\begin{equation} \label{MLE_attack_define}
\left[ {\hat \theta _{{\rm{MLE}}}^{(a)},\{ {{\hat \mu }_{\text{MLE}}^{(j)}}\} } \right]  \buildrel \Delta \over = \arg \mathop {\max }\limits_{\theta ,\{ {\mu _j} \ge b\} } {f_1}\left( {\theta ,\{ {\mu _j}\} } \right),
\end{equation}
\begin{equation} \label{MLE_unattack_define}
\hat \theta _{{\rm{MLE}}}^{(u)} \buildrel \Delta \over = \arg \mathop {\max }\limits_\theta  {f_0}\left( \theta  \right),
\end{equation}
the statistic $\Lambda _G^{(k,K)}$ can be simplified as
\begin{align} \notag
\Lambda _G^{(k,K)}  = & \sum\limits_{m = 1}^M {\sum\limits_{j \in {\cal N}} {\ln \frac{{{\bbP_0}\left( {{\check{u}}_j^{(m)}\left| {\hat \theta _{{\rm{MLE}}}^{(a)}} \right.} \right)}}{{{\bbP_0}\left( {{\check{u}}_j^{(m)}\left| {\hat \theta _{{\rm{MLE}}}^{(u)}} \right.} \right)}}} }  + \sum\limits_{i = 1}^K {\sum\limits_{j \in {\cal S}} {\ln \frac{{{\bbP_0}\left( {u_j^{(i)}\left| {\hat \theta _{{\rm{MLE}}}^{(a)}} \right.} \right)}}{{{\bbP_0}\left( {u_j^{(i)}\left| {\hat \theta _{{\rm{MLE}}}^{(u)}} \right.} \right)}}} } \\ \label{GCUSUM_statistic_simplified}
& \quad + \sum\limits_{i = 1}^{k - 1} {\sum\limits_{j \in {\cal A}} {\ln \frac{{{\bbP_0}\left( {u_j^{(i)}\left| {\hat \theta _{{\rm{MLE}}}^{(a)}} \right.} \right)}}{{{\bbP_0}\left( {u_j^{(i)}\left| {\hat \theta _{{\rm{MLE}}}^{(u)}} \right.} \right)}}} }  + \sum\limits_{i = k}^K {\sum\limits_{j \in {\cal A}} {\ln \frac{{{\bbP_1}\left( {u_j^{(i)}\left| {\hat \theta _{{\rm{MLE}}}^{(a)},{{\hat \mu }_{\text{MLE}}^{(j)}}} \right.} \right)}}{{{\bbP_0}\left( {u_j^{(i)}\left| {\hat \theta _{{\rm{MLE}}}^{(u)}} \right.} \right)}}} },
\end{align}
by employing (\ref{GCUSUM_statistic}), (\ref{MLE_attack_define}) and (\ref{MLE_unattack_define}).

It is worth mentioning that the constraint on $\{\mu_j \}$ as illustrated in (\ref{mu_constraint}) is crucial in the generalized CUSUM test in (\ref{GCUSUM_define}), since the pre-attack model will be subsumed in the post-attack model if the constraint in (\ref{mu_constraint}) is removed. To be specific, in general, in order to guarantee the detectability of the attacks, the sign of the mean value of the generalized log-likelihood ratio should differ between the pre-attack and post-attack scenarios \cite{basseville1993detection, tartakovsky2014sequential}. However, if we remove the constraint in (\ref{mu_constraint}),  we can obtain  
\begin{equation}
\Lambda _G^{(k,K)} = \mathop {\sup }\limits_{\theta ,\{ {\mu _j} \} } {f_1}\left( {\theta ,\{ {\mu _j}\} } \right) - \mathop {\sup }\limits_\theta  {f_0}\left( {\theta } \right) \ge \mathop {\sup }\limits_{\theta  } {f_1}\left( {\theta ,\{ 0\} } \right) - \mathop {\sup }\limits_\theta  {f_0}\left( {\theta } \right) = 0
\end{equation}
by employing (\ref{f1_define}) and (\ref{f0_define}). Thus, no matter whether attacks occur or not, the expected value of the test statistic is always nonnegative, which implies that the attacks are nondetectable.


\subsection{Conditions to Meet Expected False Alarm Period Constraint}

Moreover, note that the threshold $h$ in (\ref{GCUSUM_define}) needs to be set to meet the constraint on the expected false alarm period in (\ref{changepoint_detection_problem_formulation}). However, due to the uncertainty of the pre-attack model and the post-attack model, it is not clear whether the constraint in (\ref{changepoint_detection_problem_formulation}) can be guaranteed by some $h$ in general. For example, consider the case where $\hat \theta _{{\rm{MLE}}}^{(u)}$ is close to $-\infty$ with high probability. It is seen from (\ref{GCUSUM_statistic_simplified}) that 
${{{\bbP_0}( {u_j^{(i)}| {\hat \theta _{{\rm{MLE}}}^{(u)}} } )}}$ is close to $0$, and hence, $\Lambda _G^{(k,K)}$ blows up, which gives rise to $T_G$ close to $1$ with high probability for any $h$. In this case, the constraint in (\ref{changepoint_detection_problem_formulation}) may not be guaranteed. In light of this, we first look into this issue.



Notice that by defining
\begin{equation} \label{q_define}
q\left( \theta  \right) \buildrel \Delta \over = {\bbP_0}\left( {u_j^{(i)} = 0\left| \theta  \right.} \right) = F\left( {\tau  - \theta } \right),
\end{equation}
  $f_0(\theta)$ in (\ref{f0_define})   can be rewritten as
\begin{align} \notag
{f_0}\left( \theta  \right) = & \sum\limits_{m = 1}^M {\sum\limits_{j \in {\cal N}} {\left[ {\left( {1 - {\check{u}}_j^{(m)}} \right)\ln q\left( \theta  \right) + {\check{u}}_j^{(m)}\ln \left( {1 - q\left( \theta  \right)} \right)} \right]} } \\
& \quad + \sum\limits_{i = 1}^K {\sum\limits_{j \in {\cal N}} {\left[ {\left( {1 - u_j^{(i)}} \right)\ln q\left( \theta  \right) + u_j^{(i)}\ln \left( {1 - q\left( \theta  \right)} \right)} \right]} },
\end{align}
which yields
\begin{align} \label{diff_f0}
\frac{{\partial {f_0}\left( \theta  \right)}}{{\partial \theta }} = \left\{ {\sum\limits_{m = 1}^M {\sum\limits_{j \in {\cal N}} {\frac{{1 - q\left( \theta  \right) - {\check{u}}_j^{(m)}}}{{q\left( \theta  \right)\left[ {1 - q\left( \theta  \right)} \right]}}} }  + \sum\limits_{i = 1}^K {\sum\limits_{j \in {\cal N}} {\frac{{1 - q\left( \theta  \right) - u_j^{(i)}}}{{q\left( \theta  \right)\left[ {1 - q\left( \theta  \right)} \right]}}} } } \right\}\frac{{\partial q\left( \theta  \right)}}{{\partial \theta }}.
\end{align}
By the definition of $\hat \theta _{{\rm{MLE}}}^{(u)}$ in (\ref{MLE_unattack_define}), we know that $\hat \theta _{{\rm{MLE}}}^{(u)}$ is a solution   to   $\frac{{\partial {f_0}\left( \theta  \right)}}{{\partial \theta }} =0$, and hence,  we can obtain
\begin{equation} \label{theta_0_MLE}
\hat \theta _{{\rm{MLE}}}^{(u)} = \tau  - {F^{ - 1}}\left[ {1 - \frac{1}{{\left( {M + K} \right)N}}\sum\limits_{j \in {\cal N}} {\left( {\sum\limits_{m = 1}^M {{\check{u}}_j^{(m)}}  + \sum\limits_{i = 1}^K {u_j^{(i)}} } \right)} } \right]
\end{equation}
by employing (\ref{q_define}) and (\ref{diff_f0}).
In the following, we provide a theorem regarding the false alarm period.
\begin{theorem} \label{Theorem_T_G_Unattack}
	With sufficiently large $M$ or $N$, 
the false alarm period can be set to be arbitrarily large with any high probability by properly choosing $h$. In particular,
	\begin{equation}
	\mathop {\lim }\limits_{M \text{ or } N \to \infty } {\bbP_0}\left( {\left. {\mathop {\lim }\limits_{h \to \infty } {T_G} = \infty } \right|\theta } \right) = 1.
	\end{equation}
\end{theorem}
\begin{IEEEproof}
By employing (\ref{theta_0_MLE}), we can obtain
\begin{equation} \label{S_define}
F\left( {\tau  - \hat \theta _{{\rm{MLE}}}^{(u)}} \right) = 1 - \underbrace {\frac{1}{{\left( {M + K} \right)N}}\sum\limits_{j \in {\cal N}} {\left( {\sum\limits_{m = 1}^M {u_j^{(m)}}  + \sum\limits_{i = 1}^K {u_j^{(i)}} } \right)} }_{\buildrel \Delta \over = S}
\end{equation}
Denote
\begin{equation} \label{epsilon_define}
{\epsilon_1} \buildrel \Delta \over = \frac{1}{2}\left( {1 - q\left( \theta  \right)} \right) >0 \quad \text{and} \quad {\epsilon_2} \buildrel \Delta \over = \frac{1}{2}q\left( \theta  \right) >0.
\end{equation}
Now, we consider the probabilities ${\bbP_0}\left( {S \ge 1 - q\left( \theta  \right) + {\epsilon_2}} {\left| \theta  \right.}  \right)$ and ${\bbP_0}\left( {S \le 1 - q\left( \theta  \right) - {\epsilon_1}} {\left| \theta  \right.} \right)$.

Note that
\begin{align} \notag
& {\bbP_0}\left( {S \ge 1 - q\left( \theta  \right) + {\epsilon _2}\left| \theta  \right.} \right) \\ \notag
& = {\bbP_0}\left(  { \left. \frac{1}{{\left( {M + K} \right)N}}\sum\limits_{j \in {\cal N}} {\left( {\sum\limits_{m = 1}^M {\left( {{\check{u}}_j^{(m)} - 1 + q\left( \theta  \right)  } \right)}  + \sum\limits_{i = 1}^K {\left( {u_j^{(i)} - 1 + q\left( \theta  \right) } \right)} } \right)}  \ge {\epsilon _2} \right| \theta  } \right)\\ \label{prob_bound_1_1}
& = {\bbP_0}\left(  { \left. \frac{1}{{\left( {M + K} \right)N}}\sum\limits_{i = 1}^{\left( {M + K} \right)N} {{X_i}}  \ge {\epsilon _2} \right| \theta  } \right),
\end{align}
where $\{X_i\}$ is a sequence of independent and identically distributed random variables with distribution
\begin{equation} \label{pmf_X_1}
{\bbP_0}\left( {{X_i} = q\left( \theta  \right) - 1\left| \theta  \right.} \right) = q\left( \theta  \right)
\end{equation}
and
\begin{equation}  \label{pmf_X_2}
{\bbP_0}\left( {{X_i} = q\left( \theta  \right) \left| \theta  \right.} \right) = 1- q\left( \theta  \right).
\end{equation}
Since ${\epsilon _2} > \bbE_0(X_i) = 0$, by the large deviations theory \cite{dembo2010large}, we can obtain
\begin{equation} \label{prob_bound_1_2}
{\bbP_0}\left( { \left. \frac{1}{{\left( {M + K} \right)N}}\sum\limits_{i = 1}^{\left( {M + K} \right)N} {{X_i}}  \ge {\epsilon _2} \right| \theta  } \right) \le {e^{ - \upsilon _\theta ^{\left( 1 \right)}\left( {M + K} \right)N}},
\end{equation}
where the rate function $\upsilon_\theta ^{\left( 1 \right)}$ is given by
\begin{align} \notag
\upsilon _\theta ^{\left( 1 \right)} & \buildrel \Delta \over = -\mathop {\lim }\limits_{\left( {M + K} \right)N \to \infty } \frac{1}{{\left( {M + K} \right)N}}\ln {\bbP_0}\left( { \left. \sum\limits_{i = 1}^{\left( {M + K} \right)N} {{X_i}}  \ge {\epsilon _2}\left( {M + K} \right)N\right| \theta  } \right) \\ \label{rate_function_define}
& =   {\epsilon _2}{c^*} - \ln \varphi_X \left( {{c^*}} \right)
\end{align}
and the function $ \varphi_X \left( {{c}} \right)$ is defined as
\begin{equation} \label{phi_define}
{\varphi _X}\left( c \right) \buildrel \Delta \over = {\bbE_0}\left( {{e^{c{X_i}}}} \right) = q\left( \theta  \right){e^{c\left[ {q\left( \theta  \right) - 1} \right]}} + \left[1- {q\left( \theta  \right) } \right]{e^{cq\left( \theta  \right)}}.
\end{equation}
The quantity $c^*$ in (\ref{rate_function_define}) is the positive solution   to  
\begin{equation} \label{c_star_equation}
\frac{d}{{dc}}{\varphi _X}\left( c \right) + {\epsilon _2}{\varphi _X}\left( c \right) = 0.
\end{equation}
By employing (\ref{epsilon_define}), (\ref{pmf_X_1}), (\ref{pmf_X_2}), (\ref{rate_function_define}), (\ref{phi_define}) and (\ref{c_star_equation}), the rate function  $\upsilon _\theta ^{\left( 1 \right)}$ can be obtained as
\begin{align} \notag
\upsilon _\theta ^{\left( 1 \right)} & = \left[ {{\epsilon _2} - q\left( \theta  \right)} \right]\ln \frac{{q\left( \theta  \right)\left[ {1 - q\left( \theta  \right) + {\epsilon _2}} \right]}}{{\left[ {q\left( \theta  \right) - {\epsilon _2}} \right]\left[ {1 - q\left( \theta  \right)} \right]}} - \ln \frac{{1 - q\left( \theta  \right)}}{{1 - q\left( \theta  \right) + {\epsilon _2}}}\\
&  = \left[ {1 - \frac{1}{2}q\left( \theta  \right)} \right]\ln \frac{{2 - q\left( \theta  \right)}}{{1 - q\left( \theta  \right)}} - \ln 2 >0.
\end{align}
On the other hand, ${\bbP_0}\left( {S \le 1 - q\left( \theta  \right) - {\epsilon_1}} {\left| \theta  \right.} \right)$ can be rewritten as
\begin{align} \notag
& {\bbP_0}\left( {S \le 1 - q\left( \theta  \right) - {\epsilon _1}\left| \theta  \right.} \right) \\ \notag
& = {\bbP_0}\left( {\frac{1}{{\left( {M + K} \right)N}}\sum\limits_{j \in {\cal N}} {\left( {\sum\limits_{m = 1}^M {\left( {1 - q\left( \theta  \right) - \cu_j^{(m)}} \right)}  + \sum\limits_{i = 1}^K {\left( {1 - q\left( \theta  \right) - u_j^{(i)}} \right)} } \right)}  \ge {\epsilon _1}\left| \theta  \right.} \right)\\
& = {\bbP_0}\left( {\frac{1}{{\left( {M + K} \right)N}}\sum\limits_{i = 1}^{\left( {M + K} \right)N} {{Y_i}}  \ge {\epsilon _1}\left| \theta  \right.} \right),
\end{align}
where $\{Y_i\}$ is a sequence of independent and identically distributed random variables with distribution
\begin{equation} \label{pmf_Y_1}
{\bbP_0}\left( {{Y_i} = 1 - q\left( \theta  \right) \left| \theta  \right.} \right) = q\left( \theta  \right)
\end{equation}
and
\begin{equation}  \label{pmf_Y_2}
{\bbP_0}\left( {{Y_i} = -q\left( \theta  \right) \left| \theta  \right.} \right) = 1- q\left( \theta  \right).
\end{equation}
Similarly, we can obtain
\begin{equation} \label{prob_bound_2}
{\bbP_0}\left( {S \le 1 - q\left( \theta  \right) - {\epsilon _1}\left| \theta  \right.} \right) \le {e^{ - \upsilon _\theta ^{\left( 2 \right)}\left( {M + K} \right)N}},
\end{equation} 
where the rate function $\upsilon _\theta ^{\left( 2 \right)}$ is given by
\begin{equation} \label{rate_function_2}
\upsilon _\theta ^{\left( 2 \right)} = \frac{1}{2}\left[ {1 + q\left( \theta  \right)} \right]\ln \frac{{1 + q\left( \theta  \right)}}{{q\left( \theta  \right)}} - \ln 2>0.
\end{equation}

From (\ref{prob_bound_1_1}), (\ref{prob_bound_1_2}) and (\ref{prob_bound_2}), it is clear that
\begin{align} \notag
& {\bbP_0}\left( {S \in \left( {1 - q\left( \theta  \right) - {\epsilon _1},1 - q\left( \theta  \right) + {\epsilon _2}} \right)\left| \theta  \right.} \right)\\ \notag
& = 1 - {\bbP_0}\left( {S \le 1 - q\left( \theta  \right) - {\epsilon _1}\left| \theta  \right.} \right) - {\bbP_0}\left( {S \ge 1 - q\left( \theta  \right) + {\epsilon _2}\left| \theta  \right.} \right)\\ \notag
& \ge 1 - {e^{ - \upsilon _\theta ^{\left( 1 \right)}\left( {M + K} \right)N}} - {e^{ - \upsilon _\theta ^{\left( 2 \right)}\left( {M + K} \right)N}}\\
& \ge 1 - 2{e^{ - \upsilon _\theta ^*\left( {M + K} \right)N}},
\end{align}
where $\upsilon _\theta ^*$ is defined as
\begin{equation} \label{gamma_star_define}
\upsilon _\theta ^* \triangleq \min\{\upsilon _\theta ^{( 1 )}, \upsilon _\theta ^{( 2 )}\}>0.
\end{equation}

Define
\begin{equation} \label{epsilon_star_define}
{\epsilon ^*} \buildrel \Delta \over = \min \left\{ {{\epsilon _1},{\epsilon _2}} \right\} = \frac{1}{2}\min \left\{ {q\left( \theta  \right),1 - q\left( \theta  \right)} \right\} >0.
\end{equation}
Note that if ${S \in \left( {1 - q\left( \theta  \right) - {\epsilon _1},1 - q\left( \theta  \right) + {\epsilon _2}} \right)}$, then by the definition of $S$ in (\ref{S_define}), we have
\begin{equation}
F\left( {\tau  - \hat \theta _{{\rm{MLE}}}^{(u)}} \right) = 1 - S \ge q\left( \theta  \right) - {\epsilon _2} = \frac{1}{2}q\left( \theta  \right) \ge {\epsilon ^*},
\end{equation}
\begin{equation}
1 - F\left( {\tau  - \hat \theta _{{\rm{MLE}}}^{(u)}} \right) = S \ge 1 - q\left( \theta  \right) - {\epsilon _1} = \frac{1}{2}\left[ {1 - q\left( \theta  \right)} \right] \ge {\epsilon ^*},
\end{equation}
which yields
\begin{equation} \label{F_min_Define}
{F_{\min }} \buildrel \Delta \over = \min \left\{ {F\left( {\tau  - \hat \theta _{{\rm{MLE}}}^{(u)}} \right),1 - F\left( {\tau  - \hat \theta _{{\rm{MLE}}}^{(u)}} \right)} \right\} \ge {\epsilon ^*}.
\end{equation}
Moreover, if  ${S \in \left( {1 - q\left( \theta  \right) - {\epsilon _1},1 - q\left( \theta  \right) + {\epsilon _2}} \right)}$, then by employing (\ref{F_min_Define}), we can obtain
\begin{equation}
{\bbP_0}\left( {u\left| {\hat \theta _{{\rm{MLE}}}^{(u)}} \right.} \right) = \left( {1 - u} \right)F\left( {\tau  - \hat \theta _{{\rm{MLE}}}^{(u)}} \right) + u\left[ {1 - F\left( {\tau  - \hat \theta _{{\rm{MLE}}}^{(u)}} \right)} \right] \ge {F_{\min }} \ge {\epsilon ^*} > 0, \; \forall u=0,1,
\end{equation}
which implies that for any $\cu_j^{(m)}$, $u_j^{(i)}$, ${\hat \theta _{{\rm{MLE}}}^{(u)}}$, ${\hat \theta _{{\rm{MLE}}}^{(a)}}$ and ${\hat \mu }_{\text{MLE}}^{(j)}$, 
\begin{equation}
\ln \frac{{{\bbP_0}\left( {\cu_j^{(m)}\left| {\hat \theta _{{\rm{MLE}}}^{(a)}} \right.} \right)}}{{{\bbP_0}\left( {\cu_j^{(m)}\left| {\hat \theta _{{\rm{MLE}}}^{(u)}} \right.} \right)}}, \;\; \ln \frac{{{\bbP_0}\left( {u_j^{(i)}\left| {\hat \theta _{{\rm{MLE}}}^{(a)}} \right.} \right)}}{{{\bbP_0}\left( {u_j^{(i)}\left| {\hat \theta _{{\rm{MLE}}}^{(u)}} \right.} \right)}}, \;\;\ln \frac{{{\bbP_1}\left( {u_j^{(i)}\left| {\hat \theta _{{\rm{MLE}}}^{(a)},{{\hat \mu }_{\text{MLE}}^{(j)}}} \right.} \right)}}{{{\bbP_0}\left( {u_j^{(i)}\left| {\hat \theta _{{\rm{MLE}}}^{(u)}} \right.} \right)}} \le  - \ln {\epsilon ^*},
\end{equation}
and therefore, by employing (\ref{GCUSUM_statistic_simplified}), we can obtain
\begin{equation} \label{Lambda_ub_1}
\Lambda _G^{(k,K)} \le  - \left( {M + K} \right)N\ln {\epsilon ^*}.
\end{equation}

Note that for any given $M$, $N$ and $h$, if (\ref{Lambda_ub_1}) holds and the stopping criterion in (\ref{GCUSUM_define}) is triggered, then $T_G = K$ and
\begin{equation}
- \left( {M + T_G} \right)N\ln {\epsilon ^*} \ge {\mathop {\max }\limits_{1 \le k \le K} \Lambda _G^{(k,K)}} \ge h,
\end{equation}
which implies
\begin{equation}
{T_G} \ge  - \frac{h}{{N\ln {\epsilon ^*}}} - M.
\end{equation}
In consequence, for any given $M$, $N$ and $h$, we have
\begin{align} \notag
& {\bbP_0}\left( { \left. {T_G} \ge  - \frac{h}{{N\ln {\epsilon ^*}}} - M \right| \theta  } \right) \\ \notag
&  \ge {\bbP_0}\left( {S \in \left( {1 - q\left( \theta  \right) - {\epsilon _1},1 - q\left( \theta  \right) + {\epsilon _2}} \right)\left| \theta  \right.} \right)\\ \notag
& \ge 1 - 2{e^{ - \upsilon _\theta ^*\left( {M + {T_G}} \right)N}}\\ \label{T_G_larger_prob}
& \ge 1 - 2{e^{ - \upsilon _\theta ^*MN}},
\end{align}
which yields that for any given $\theta$,
\begin{equation}
\mathop {\lim }\limits_{M \text{ or } N \to \infty } {\bbP_0}\left( {\left. {\mathop {\lim }\limits_{h \to \infty } {T_G} = \infty } \right|\theta } \right) = 1.
\end{equation}
From (\ref{T_G_larger_prob}), we know that with sufficiently large $M$ or $N$, if there is no occurrence of attacks, the stopping time $T_G$ in (\ref{changepoint_detection_problem_formulation}) can be set to be arbitrarily large with any high probability by properly choosing $h$.
\end{IEEEproof}

Next, we provide another theorem regarding the constraint on the expected false alarm period in (\ref{changepoint_detection_problem_formulation}).
\begin{theorem} \label{Theorem_False_Alarm_Peroid}
	If $MN$ is large enough such that
	\begin{equation} \label{MN_constraint_corollary}
	MN > \frac{{\ln 2}}{{\upsilon _\theta ^*}}
	\end{equation}
	where ${\upsilon _\theta ^*}$ is defined in (\ref{gamma_star_define}), then ${\bbE_\infty }\left\{ {{T_G}} \right\} \ge \kappa$ as long as
	\begin{equation} \label{h_condition}
	h \ge N\left( {\frac{\kappa }{{1 - 2{e^{ - \upsilon _\theta ^*MN}}}} + M} \right)\ln \frac{1}{{{\epsilon ^*}}},
	\end{equation}
	where $\epsilon ^*$ is defined in (\ref{epsilon_star_define}).
\end{theorem}
\begin{IEEEproof}
	If $MN > \frac{{\ln 2}}{{\upsilon _\theta ^*}}$, then we know that 
	\begin{equation}
	{1 - 2{e^{ - \upsilon _\theta ^*MN}}} > 0.
	\end{equation}
Since $T_G$ is a non-negative random variable, the expected false alarm period can be bounded from below as per

\begin{align} \notag
{\bbE_\infty }\left\{ {{T_G}} \right\} & = \int_0^\infty  {{\bbP_0}\left( {{T_G} \ge x\left| \theta  \right.} \right)dx} \\ \notag
& \ge \int_0^{ - \frac{h}{{N\ln {\epsilon ^*}}} - M} {{\bbP_0}\left( {{T_G} \ge x\left| \theta  \right.} \right)dx} \\ \notag
& \ge {\bbP_0}\left( { \left. {T_G} \ge  - \frac{h}{{N\ln {\epsilon ^*}}} - M \right| \theta  } \right)\int_0^{ - \frac{h}{{N\ln {\epsilon ^*}}} - M} {dx} \\ \notag
& \ge \left( {1 - 2{e^{ - \upsilon _\theta ^*MN}}} \right)\left( { - \frac{h}{{N\ln {\epsilon ^*}}} - M} \right) \\
& \ge \kappa
\end{align}
by employing (\ref{T_G_larger_prob}) and (\ref{h_condition}), which completes the proof.
\end{IEEEproof}

As demonstrated by {Theorem \ref{Theorem_T_G_Unattack}}, in the considered attack detection problem, if the length of secure time $M$ or the number of sensors $N$ is large enough, the false alarm period can be guaranteed to be arbitrarily large with any high probability by choosing a large $h$. A larger false alarm period is always desirable, hence the system would better maintain secure for a sufficiently large $M$ prior to the inference task.   Moreover,   {Theorem \ref{Theorem_False_Alarm_Peroid}} illustrates that in order to guarantee the expected false alarm period to be larger than the prescribed threshold $\kappa$ in (\ref{changepoint_detection_problem_formulation}), the product of $M$ and $N$ should be larger than the critical quantity $\frac{{\ln 2}}{{\upsilon _\theta ^*}}$. However, as shown in (\ref{rate_function_define}), (\ref{rate_function_2}) and (\ref{gamma_star_define}), this critical quantity depends on $\theta$ which is unknown to the system, and therefore, to guarantee the condition in (\ref{MN_constraint_corollary}), the system should be secured for a sufficiently long time in advance, i.e., $M$ should be sufficiently large. In light of these, in the following, we will be primarily interested in the case where $M$ is sufficiently large.

\subsection{Approximate Maximum Likelihood Estimates}

By employing (\ref{f1_define}), (\ref{MLE_attack_define}) and the Karush-Kuhn-Tucker conditions \cite{boyd2004convex}, $\hat \theta _{{\rm{MLE}}}^{(a)}$ and  ${{\hat \mu }_{\text{MLE}}^{(j)}} $ are the solutions   to    
\begin{numcases}{} \label{eq_1}
 \frac{{\partial {f_1}\left( {\theta ,\{ {\mu _j}\} } \right)}}{{\partial \theta }} = 0, \\ \label{eq_2}
 \frac{{\partial {f_1}\left( {\theta ,\{ {\mu _j}\} } \right)}}{{\partial {\mu _j}}} + {c_j} = 0, \\ \label{eq_3}
 {c _j}\left( {b - {\mu _j}} \right) = 0,  \\ \label{eq_4}
 {c _j} \ge 0,  \\ \label{eq_5}
 {\mu _j} \ge b,
\end{numcases} 
where ${f_1}\left( {\theta ,\{ {\mu _j}\} } \right)$ is defined in (\ref{f1_define}). By defining
\begin{equation} \label{tilde_q_define}
\tilde q\left( {\theta ,{\mu _j}} \right) \buildrel \Delta \over = F\left( {\tau  - \theta  - {\mu _j}} \right),
\end{equation}
$\frac{{\partial {f_1}\left( {\theta ,\{ {\mu _j}\} } \right)}}{{\partial \mu_j }}$ and $\frac{{\partial {f_1}\left( {\theta ,\{ {\mu _j}\} } \right)}}{{\partial \theta }}$ can be analytically expressed as
\begin{equation} \label{df1_dmu}
\frac{{\partial {f_1}\left( {\theta ,\{ {\mu _j}\} } \right)}}{{\partial {\mu _j}}} = \frac{{\partial \tilde q\left( {\theta ,{\mu _j}} \right)}}{{\partial {\mu _j}}}\sum\limits_{i = k}^K {\frac{{1 - \tilde q\left( {\theta ,{\mu _j}} \right) - u_j^{\left( i \right)}}}{{\tilde q\left( {\theta ,{\mu _j}} \right)\left[ {1 - \tilde q\left( {\theta ,{\mu _j}} \right)} \right]}}}, 
\end{equation}
\begin{align} \notag
\frac{{\partial {f_1}\left( {\theta ,\{ {\mu _j}\} } \right)}}{{\partial \theta }}&  = \frac{{\partial q\left( \theta  \right)}}{{\partial \theta }}\Bigg\{ \sum\limits_{m = 1}^M {\sum\limits_{j \in {\cal N}} {\frac{{1 - q\left( \theta  \right) - \cu_j^{(m)}}}{{q\left( \theta  \right)\left[ {1 - q\left( \theta  \right)} \right]}}} }  + \sum\limits_{i = 1}^K {\sum\limits_{j \in {\cal S}} {\frac{{1 - q\left( \theta  \right) - u_j^{(i)}}}{{q\left( \theta  \right)\left[ {1 - q\left( \theta  \right)} \right]}}} } \\ \label{df1_dtheta}
& \quad \quad + \sum\limits_{i = 1}^{k - 1} {\sum\limits_{j \in {\cal A}} {\frac{{1 - q\left( \theta  \right) - u_j^{(i)}}}{{q\left( \theta  \right)\left[ {1 - q\left( \theta  \right)} \right]}}} }  \Bigg\}   + \frac{{\partial \tilde q\left( {\theta ,{\mu _j}} \right)}}{{\partial \theta }}\sum\limits_{i = k}^K {\sum\limits_{j \in {\cal A}} {\frac{{1 - \tilde q\left( {\theta ,{\mu _j}} \right) - u_j^{(i)}}}{{\tilde q\left( {\theta ,{\mu _j}} \right)\left[ {1 - \tilde q\left( {\theta ,{\mu _j}} \right)} \right]}}} },
\end{align}
where $q( \theta  )$ is defined in (\ref{q_define}). By employing (\ref{eq_2})--(\ref{eq_5}) and (\ref{df1_dmu}), we can obtain
\begin{equation} \label{mu_MLE_constraint}
{{\hat \mu }_{\text{MLE}}^{(j)}} = \max \left\{ {{{\tilde \mu }_j},b} \right\},
\end{equation}
where ${\tilde \mu }_j$ is defined as
\begin{equation} \label{mu_MLE_unconstraint}
{{\tilde \mu }_j} \buildrel \Delta \over = \tau  - \hat \theta _{{\rm{MLE}}}^{(a)} - {F^{ - 1}}\left[ {1 - \frac{1}{{K - k + 1}}\sum\limits_{i = k}^K {u_j^{(i)}} } \right].
\end{equation}
It is worth mentioning that ${{\hat \mu }_{\text{MLE}}^{(j)}}$ and $\hat \theta _{{\rm{MLE}}}^{(a)}$ can only be numerically obtained by using (\ref{eq_1}), (\ref{df1_dtheta}), (\ref{mu_MLE_constraint}) and (\ref{mu_MLE_unconstraint}). 
Next, we propose to employ the closed-form approximations to ${{\hat \mu }_{\text{MLE}}^{(j)}}$ and $\hat \theta _{{\rm{MLE}}}^{(a)}$. 
In particular, we employ ${\hat \theta}^{(a)}$ as a substitute for $\hat \theta _{{\rm{MLE}}}^{(a)}$, which is defined as
\begin{equation} \label{theta_est}
{{\hat \theta }^{(a)}} \buildrel \Delta \over = \tau   - {F^{ - 1}}\left[ {1 - \frac{1}{{MN + K{N_{\cal S}}}}\left( {\sum\limits_{j \in {\cal N}} {\sum\limits_{m = 1}^M {\cu_j^{(m)}} }  + \sum\limits_{j \in {\cal S}} {\sum\limits_{i = 1}^K {u_j^{(i)}} } } \right)} \right].
\end{equation}

The motivation for ${{\hat \theta }^{(a)}}$ in (\ref{theta_est}) is from the fact that in the asymptotic regimes where $M \to \infty$ or $K \to \infty$, from (\ref{theta_est}), we know
\begin{equation}
{{\hat \theta }^{(a)}} \to \theta \quad \text{almost surely (a.s.)},
\end{equation}
by the strong law of large numbers.
As stated before, we are primarily interested in the scenario where the system is secured for a sufficiently long time at the beginning, i.e., $M$ is sufficiently large, thus the estimate ${{\hat \theta }^{(a)}}$ converges to the true value of the unknown parameter in the scenario of interest. Moreover, with a sufficiently large number of observations, $\hat \theta _{{\rm{MLE}}}^{(a)}$ is also expected to get close to the true value. To this end, ${{\hat \theta }^{(a)}}$  must be close to $\hat \theta _{{\rm{MLE}}}^{(a)}$ in the scenario of interest, and hence, is a good substitute for $\hat \theta _{{\rm{MLE}}}^{(a)}$ which cannot be analytically expressed.

By employing (\ref{mu_MLE_unconstraint}) and (\ref{theta_est}), the corresponding ${\tilde \mu }_j$ can be expressed as
\begin{equation} \label{tilde_mu_approx}
{{\tilde \mu }_j}  \approx {F^{ - 1}}\left[ {1 \! - \frac{1}{{MN + K{N_{\cal S}}}}\left( {\sum\limits_{l \in {\cal N}} {\sum\limits_{m = 1}^M {\cu_l^{(m)}} }  + \sum\limits_{l \in {\cal S}} {\sum\limits_{i = 1}^K {u_l^{(i)}} } } \right)} \right]  -   {F^{ - 1}}\left[ {1 \!  - \frac{1}{{K - k + 1}}\sum\limits_{i = k}^K {u_j^{(i)}} } \right],
\end{equation}
and $\Lambda _G^{(k,K)}$ can be rewritten as
\begin{align} \notag 
\Lambda _G^{(k,K)} & \approx \sum\limits_{m = 1}^M {\sum\limits_{j \in {\cal N}} {\underbrace {\ln \frac{{{\bbP_0}\left( {\cu_j^{(m)}\left| {{{\hat \theta }^{(a)}}} \right.} \right)}}{{{\bbP_0}\left( {\cu_j^{(m)}\left| {\hat \theta _{{\rm{MLE}}}^{(u)}} \right.} \right)}}}_{ \buildrel \Delta \over = \phi _{jm}^{\left( 1 \right)}}} }  + \sum\limits_{i = 1}^K {\sum\limits_{j \in {\cal S}} {\underbrace {\ln \frac{{{\bbP_0}\left( {u_j^{(i)}\left| {{{\hat \theta }^{(a)}}} \right.} \right)}}{{{\bbP_0}\left( {u_j^{(i)}\left| {\hat \theta _{{\rm{MLE}}}^{(u)}} \right.} \right)}}}_{ \buildrel \Delta \over = \phi _{ji}^{\left( 2 \right)}}} } \\ \label{Lambda_G_Simplified}
& \qquad + \sum\limits_{i = 1}^{k - 1} {\sum\limits_{j \in {\cal A}} {\underbrace {\ln \frac{{{\bbP_0}\left( {u_j^{(i)}\left| {{{\hat \theta }^{(a)}}} \right.} \right)}}{{{\bbP_0}\left( {u_j^{(i)}\left| {\hat \theta _{{\rm{MLE}}}^{(u)}} \right.} \right)}}}_{ \buildrel \Delta \over = \phi _{ji}^{\left( 3 \right)}}} }  + \sum\limits_{i = k}^K {\sum\limits_{j \in {\cal A}} {\underbrace {\ln \frac{{{\bbP_1}\left( {u_j^{(i)}\left| {{{\hat \theta }^{(a)}},{{\hat \mu }_j}} \right.} \right)}}{{{\bbP_0}\left( {u_j^{(i)}\left| {\hat \theta _{{\rm{MLE}}}^{(u)}} \right.} \right)}}}_{ \buildrel \Delta \over = \phi _{ji}^{\left( 4 \right)}}} },
\end{align}
where ${\hat \theta _{{\rm{MLE}}}^{(u)}}$ is given by (\ref{theta_0_MLE}) and ${{\hat \mu }_j} \buildrel \Delta \over = \max \left\{ {{{\tilde \mu }_j},b} \right\}$.




It is worth pointing out that the estimators  ${{{\hat \theta }^{(a)}}}$, ${\hat \theta _{{\rm{MLE}}}^{(u)}}$ and ${{\hat \mu }_j}$ need to be recalculated for different $k$ or $K$, which results in a very high computational complexity in obtaining $T_G$ in (\ref{GCUSUM_define}). 
  Moreover, it is seen from (\ref{GCUSUM_define}) that  since the test statistic of generalized CUSUM can not be calculated in a recursive way, the computation of the test statistic of the generalized CUSUM requires an exhaustive search over $k$ for each $K$ which is considered as computationally prohibitive in practice, especially when $h$ is large.
To this end,  the generalized CUSUM test in (\ref{GCUSUM_define}) is not amenable to implementation in practice even if  centralized processing is allowed, which motives us to pursue more practical algorithms in the following.

The test statistic of the generalized CUSUM in (\ref{Lambda_G_Simplified}) deserves some discussion. 
It is seen from (\ref{Lambda_G_Simplified}) that the first two terms in the sum coming from the secure data which is due to the existence of unknown parameters in  both the pre-attack and post-attack models. This implies that in the presence of the unknown parameters in  both the pre-attack and post-attack models, the secure data also conveys some information on the attacks, which is not observed in the test statistic of CUSUM in (\ref{cusum_test}).
In addition, if $\theta$ is known, then it is clear that the first three terms in (\ref{Lambda_G_Simplified}) varnish, which greatly simplifies the expression of the test statistic in (\ref{Lambda_G_Simplified}). Therefore, the presence of unknown parameters in  both the pre-attack and post-attack models renders the attack detection problem more difficult.

\section{Consensus-based Distributed Sequential Attack Detection}
\label{Section_DCUSUM}

In the previous section, the generalized CUSUM detector is investigated. It is seen from (\ref{GCUSUM_define}), (\ref{theta_0_MLE}), (\ref{mu_MLE_constraint}), (\ref{theta_est}) and (\ref{Lambda_G_Simplified}) that all the quantized sensor data in the network is required to be synchronously collected and processed at each time at a single place which plays the role of the fusion center. This brings about a significant communication overhead in large-scale spatially distributed sensor networks.   More importantly, as mentioned before, the generalized CUSUM is computationally prohibitive in practice.  
In light of this, we proceed to propose a distributed sequential attack detector based on the generalized CUSUM test,   which only requires  communications between closely positioned sensors and is  amenable to implementation in practice.   The resulting distributed detector remarkably decreases the long-distance communication overhead   and reduces the computational complexity,    yet preserves high detection performance.  

\subsection{Asymptotically Equivalent Alternative Test Statistic}

For each given $K$, the test statistic $ {\max }_{1 \le k \le K} \Lambda _G^{(k,K)}$ of the generalized CUSUM test can be rewritten as
\begin{equation} \label{H_G_Define}
H_{K,M}^{\left( G \right)} \buildrel \Delta \over = \mathop {\max }\limits_{1 \le k \le K} \Lambda _G^{(k,K)} = \underbrace {\sum\limits_{m = 1}^M {\sum\limits_{j \in {\cal N}} {\phi _{jm}^{\left( 1 \right)}} } }_{ \buildrel \Delta \over = \eta _{K,M}^{(1)}} + \underbrace {\sum\limits_{i = 1}^K {\sum\limits_{j \in {\cal S}} {\phi _{ji}^{\left( 2 \right)}} } }_{ \buildrel \Delta \over = \eta _{K,M}^{(2)}} + \mathop {\max }\limits_{1 \le k \le K} \Bigg\{ {\underbrace {\sum\limits_{i = 1}^{k - 1} {\sum\limits_{j \in {\cal A}} {\phi _{ji}^{\left( 3 \right)}} } }_{ \buildrel \Delta \over = \eta _{K,M}^{(3)}} + \underbrace {\sum\limits_{i = k}^K {\sum\limits_{j \in {\cal A}} {\phi _{ji}^{\left( 4 \right)}} } }_{ \buildrel \Delta \over = \eta _{K,M}^{(4)}}} \Bigg\}
\end{equation}
by employing (\ref{GCUSUM_define}) and (\ref{Lambda_G_Simplified}).   In this subsection, in order to facilitate the distributed implementation of the generalized CUSUM test in (\ref{GCUSUM_define}) later, we first propose an alternative test statistic as a substitute for $H_{K,M}^{\left( G \right)}$, which is equivalent to $H_{K,M}^{\left( G \right)}$ in the asymptotic regime of interest where $M$ is sufficiently large.  

It is worth mentioning that although the term $\eta _{K,M}^{(3)}$ in (\ref{H_G_Define}) is the log-likelihood ratio of the quantized data from $\cA$ up to time $k \le K$, it is adapted to the filtration $\cF_K$ generated by all quantized data from all sensors up to time $K$, since $k$ is adapted to the filtration $\cF_K$. Therefore, the computation of  $\eta _{K,M}^{(3)}$ requires all  quantized data from all sensors up to time $K$, which hinders us from  computing $H_{K,M}^{\left( G \right)}$ in a distributed and efficient way. In light of this, a straightforward idea is to get rid of $\eta _{K,M}^{(3)}$ in (\ref{H_G_Define}) to facilitate the distributed computation of the test statistic, which yields an alternative test statistic given by
\begin{equation} \label{H_A_Define}
H_{K,M}^{\left( A \right)} \buildrel \Delta \over =  \eta _{K,M}^{(1)} + \eta _{K,M}^{(2)} + \mathop {\max }\limits_{1 \le k \le K}  \eta _{K,M}^{(4)}.
\end{equation}


The following theorem describes the relationship between $H_{K,M}^{\left( A \right)}$ and $H_{K,M}^{\left( G \right)}$.

\begin{theorem} \label{Theorem_H_A_H_G}
	In the asymptotic regime where   $M \to \infty$,   for any given $K<\infty$, we have
	\begin{equation}
	\mathop {\lim }\limits_{M \to \infty } \left| {H_{K,M}^{\left( A \right)} - H_{K,M}^{\left( G \right)}} \right| = 0 \quad \text{a.s.}
	\end{equation}
	no matter whether attacks occur or not.
\end{theorem}
\begin{IEEEproof}
From (\ref{H_G_Define}) and (\ref{H_A_Define}), we can obtain
\begin{equation} \label{H_limit_diff_1}
\mathop {\lim }\limits_{M \to \infty } \left| {H_{K,M}^{\left( A \right)} - H_{K,M}^{\left( G \right)}} \right|  = \left| {\mathop {\lim }\limits_{M \to \infty } \mathop {\max }\limits_{1 \le k \le K} \left\{ {\eta _{K,M}^{(3)} + \eta _{K,M}^{(4)}} \right\} -  \mathop {\lim }\limits_{M \to \infty } \mathop {\max }\limits_{1 \le k \le K}  \eta _{K,M}^{(4)}} \right|.
\end{equation}
First, we claim that we can switch the orders of the limit operation and the max operation for the first term on the right-hand side in (\ref{H_limit_diff_1}). This claim can be proved as follows. 

For any two random variables $X_M$ and $Y_M$, 
\begin{equation}
\max \left\{ {{X_M},{Y_M}} \right\} = \frac{1}{2}\left( {{X_M} + {Y_M} + \left| {{X_M} - {Y_M}} \right|} \right)
\end{equation}
which is a continuous function of $X_M$ and $Y_M$, and therefore,
\begin{equation}
\mathop {\lim }\limits_{M \to \infty } \max \left\{ {{X_M},{Y_M}} \right\} = \max \left\{ {\mathop {\lim }\limits_{M \to \infty } {X_M},\mathop {\lim }\limits_{M \to \infty } {Y_M}} \right\}.
\end{equation}
By induction, for any given $K<\infty$, we know
\begin{equation}
\mathop {\lim }\limits_{M \to \infty } \mathop {\max }\limits_{1 \le k \le K} \left\{ {\eta _{K,M}^{(3)} + \eta _{K,M}^{(4)}} \right\} = \mathop {\max }\limits_{1 \le k \le K} \left\{ {\mathop {\lim }\limits_{M \to \infty } \eta _{K,M}^{(3)} + \mathop {\lim }\limits_{M \to \infty } \eta _{K,M}^{(4)}} \right\}
\end{equation}
which yields
\begin{equation} \label{H_limit_diff_2}
\mathop {\lim }\limits_{M \to \infty } \left| {H_{K,M}^{\left( A \right)} -  H_{K,M}^{\left( G \right)}} \right|  =  \left| {\mathop {\max }\limits_{1 \le k \le K} \left\{ {\mathop {\lim }\limits_{M \to \infty } \eta _{K,M}^{(3)} + \mathop {\lim }\limits_{M \to \infty } \eta _{K,M}^{(4)}} \right\} - \mathop {\lim }\limits_{M \to \infty } \mathop {\max }\limits_{1 \le k \le K}  \eta _{K,M}^{(4)}} \right|.
\end{equation}

By employing the definition of ${\phi _{ji}^{\left( 3 \right)}}$ in (\ref{Lambda_G_Simplified}), we can obtain
\begin{equation}
\phi _{ji}^{\left( 3 \right)} = \ln \frac{{{\bbP_0}\left( {u_j^{(i)}\left| {{{\hat \theta }^{(a)}}} \right.} \right)}}{{{\bbP_0}\left( {u_j^{(i)}\left| {\hat \theta _{{\rm{MLE}}}^{(u)}} \right.} \right)}} = \left( {1 - u_j^{(i)}} \right)\ln \frac{{F\left( {\tau  - {{\hat \theta }^{\left( a \right)}}} \right)}}{{F\left( {\tau  - \hat \theta _{{\rm{MLE}}}^{(u)}} \right)}} + u_j^{(i)}\ln \frac{{1 - F\left( {\tau  - {{\hat \theta }^{\left( a \right)}}} \right)}}{{1 - F\left( {\tau  - \hat \theta _{{\rm{MLE}}}^{(u)}} \right)}},
\end{equation}
which implies
\begin{align} \notag
\left| {\eta _{K,M}^{(3)}} \right| & = \left| {\sum\limits_{i = 1}^{k - 1} {\sum\limits_{j \in {\cal A}} {\phi _{ji}^{\left( 3 \right)}} } } \right| \\ \notag
& \le \left( {k - 1} \right){N_{\cal A}}\left[ {\left| {\ln \frac{{F\left( {\tau  - {{\hat \theta }^{\left( a \right)}}} \right)}}{{F\left( {\tau  - \hat \theta _{{\rm{MLE}}}^{(u)}} \right)}}} \right| + \left| {\ln \frac{{1 - F\left( {\tau  - {{\hat \theta }^{\left( a \right)}}} \right)}}{{1 - F\left( {\tau  - \hat \theta _{{\rm{MLE}}}^{(u)}} \right)}}} \right|} \right]\\ \label{eta_3_ub}
& \le KN\left[ {\left| {\ln \frac{{F\left( {\tau  - {{\hat \theta }^{\left( a \right)}}} \right)}}{{F\left( {\tau  - \hat \theta _{{\rm{MLE}}}^{(u)}} \right)}}} \right| + \left| {\ln \frac{{1 - F\left( {\tau  - {{\hat \theta }^{\left( a \right)}}} \right)}}{{1 - F\left( {\tau  - \hat \theta _{{\rm{MLE}}}^{(u)}} \right)}}} \right|} \right].
\end{align}
Furthermore, notice that for any given $K \le \infty$, 
\begin{align} \notag
\mathop {\lim }\limits_{M \to \infty } \ln \frac{{F\left( {\tau  - {{\hat \theta }^{\left( a \right)}}} \right)}}{{F\left( {\tau  - \hat \theta _{{\rm{MLE}}}^{(u)}} \right)}} & = \ln \frac{{1 - \mathop {\lim }\limits_{M \to \infty } \frac{1}{{MN + K{N_{\cal S}}}}\left( {\sum\limits_{j \in {\cal N}} {\sum\limits_{m = 1}^M {\cu_j^{(m)}} }  + \sum\limits_{j \in {\cal S}} {\sum\limits_{i = 1}^K {u_j^{(i)}} } } \right)}}{{1 - \mathop {\lim }\limits_{M \to \infty } \frac{1}{{\left( {M + K} \right)N}}\sum\limits_{j \in {\cal N}} {\left( {\sum\limits_{m = 1}^M {u_j^{(m)}}  + \sum\limits_{i = 1}^K {u_j^{(i)}} } \right)} }}\\ \label{explain_1}
& = \ln \frac{{F\left( {\tau  - \theta } \right)}}{{F\left( {\tau  - \theta } \right)}}, \;\; \text{a.s.}\\ \label{lnF_temp_1}
& = 0,
\end{align}
\begin{align} \notag
\mathop {\lim }\limits_{M \to \infty } \ln \frac{{1 - F\left( {\tau  - {{\hat \theta }^{\left( a \right)}}} \right)}}{{1 - F\left( {\tau  - \hat \theta _{{\rm{MLE}}}^{(u)}} \right)}} & = \ln \frac{{\mathop {\lim }\limits_{M \to \infty } \frac{1}{{MN + K{N_{\cal S}}}}\left( {\sum\limits_{j \in {\cal N}} {\sum\limits_{m = 1}^M {\cu_j^{(m)}} }  + \sum\limits_{j \in {\cal S}} {\sum\limits_{i = 1}^K {u_j^{(i)}} } } \right)}}{{\mathop {\lim }\limits_{M \to \infty } \frac{1}{{\left( {M + K} \right)N}}\sum\limits_{j \in {\cal N}} {\left( {\sum\limits_{m = 1}^M {u_j^{(m)}}  + \sum\limits_{i = 1}^K {u_j^{(i)}} } \right)} }}\\ \label{explain_2}
& = \ln \frac{{1 - F\left( {\tau  - \theta } \right)}}{{1 - F\left( {\tau  - \theta } \right)}}, \;\; \text{ a.s.}\\ \label{lnF_temp_2}
& = 0,
\end{align}
where (\ref{explain_1}) and (\ref{explain_2}) are from the strong law of large numbers that
\begin{equation}
\mathop {\lim }\limits_{M \to \infty } \frac{1}{{MN}}\sum\limits_{j \in {\cal N}} {\sum\limits_{m = 1}^M {\cu_j^{(m)}} }  = 1 - F\left( {\tau  - \theta } \right), \;\; \text{ a.s.}.
\end{equation}
By employing (\ref{eta_3_ub}), (\ref{lnF_temp_1}) and (\ref{lnF_temp_2}), we can obtain
\begin{equation}
\mathop {\lim }\limits_{M \to \infty } \left| {\eta _{K,M}^{(3)}} \right|=0,  \;\; \text{ a.s.}.
\end{equation}
As a result, from (\ref{H_limit_diff_2}), we know that for any given $K \le \infty$
\begin{align} \notag
\mathop {\lim }\limits_{M \to \infty } \left| {H_{K,M}^{\left( A \right)} - H_{K,M}^{\left( G \right)}} \right| & = \left| {\mathop {\max }\limits_{1 \le k \le K} \left\{ {\mathop {\lim }\limits_{M \to \infty } \eta _{K,M}^{(3)} + \mathop {\lim }\limits_{M \to \infty } \eta _{K,M}^{(4)}} \right\} - \mathop {\lim }\limits_{M \to \infty } \mathop {\max }\limits_{1 \le k \le K}  \eta _{K,M}^{(4)}} \right|\\ \notag
& = \left| {\mathop {\max }\limits_{1 \le k \le K} \mathop {\lim }\limits_{M \to \infty } \eta _{K,M}^{(4)} - \mathop {\lim }\limits_{M \to \infty } \mathop {\max }\limits_{1 \le k \le K}  \eta _{K,M}^{(4)}} \right|,  \;\; \text{ a.s.}\\  \notag
& = \left|  \mathop {\lim }\limits_{M \to \infty } {\mathop {\max }\limits_{1 \le k \le K} \eta _{K,M}^{(4)} - \mathop {\lim }\limits_{M \to \infty } \mathop {\max }\limits_{1 \le k \le K}  \eta _{K,M}^{(4)}} \right| \\
& = 0,
\end{align}
which completes the proof.
\end{IEEEproof}

As demonstrated by \emph{Theorem \ref{Theorem_H_A_H_G}}, if $K$ is finite, then the alternative test statistic $H_{K,M}^{\left( A \right)}$ is asymptotically equivalent to the test statistic $H_{K,M}^{\left( G \right)}$ when $M$ is sufficiently large. It is worth mentioning that in general, if $h$ in (\ref{GCUSUM_define}) is finite, then $T_G<\infty$ with probability one, which implies $K<\infty$ since $K\le T_G$. To this end, in the asymptotic regime where $M \to \infty$, we can employ the following test instead of the generalized CUSUM test in (\ref{GCUSUM_define})
\begin{equation} \label{Test_Alternative}
{T_A} = \min \left\{ {K:H_{K,M}^{(A)} \ge h} \right\}.
\end{equation}

Noticing from (\ref{H_A_Define}), the test statistic $H_{K,M}^{\left( A \right)}$ consists of three terms, that is, ${{\eta} _{K,M}^{(1)}}$, ${{\eta} _{K,M}^{(2)}}$ and $\max_{1 \le k \le K} {{\eta} _{K,M}^{(4)}}$. 
The terms ${{\eta} _{K,M}^{(1)}}$ and ${{\eta} _{K,M}^{(2)}}$ come from the information on attacks conveyed in the secure data, while $\max_{1 \le k \le K} {{\eta} _{K,M}^{(4)}}$ is contributed by the insecure data.
In the following subsections, we will illustrate how to compute these three terms in a distributed way by employing running consensus algorithms, respectively.

\subsection{Local Estimators Based on Running Consensus Algorithms}


In this subsection, we  consider the distributed computation of certain statistics based on the communication protocol known as the running consensus algorithm, in which the sensors exchange their local statistics instead of the raw samples between every two sampling instants. Moreover, we assume that $Q \ge 1$ rounds of message-passings can take place within each sampling interval. 

Let us first consider the statistics in $\eta _{K,M}^{(1)}$ in (\ref{H_G_Define}).
From (\ref{theta_0_MLE}), (\ref{theta_est}), (\ref{Lambda_G_Simplified}) and (\ref{H_G_Define}), $\eta _{K,M}^{(1)}$ can be rewritten as
\begin{equation} \label{eta_K_M_1_rewritten}
\eta _{K,M}^{(1)} = \left( {MN - {\lambda _{\cal M}}} \right)G_{K,M}^{(1)} + {\lambda _{\cal M}}G_{K,M}^{(2)},
\end{equation}
where 
\begin{equation} \label{G_define}
G_{K,M}^{(1)} \buildrel \Delta \over = \ln \frac{{1 - \frac{1}{{MN + K{N_{\cal S}}}}\left( {{\lambda _{\cal M}} + {\lambda _{\cal S}}} \right)}}{{1 - \frac{1}{{MN + KN}}\left( {{\lambda _{\cal M}} + {\lambda _{\cal N}}} \right)}}, \;\; G_{K,M}^{(2)} \buildrel \Delta \over = \ln \frac{{\frac{1}{{MN + K{N_{\cal S}}}}\left( {{\lambda _{\cal M}} + {\lambda _{\cal S}}} \right)}}{{\frac{1}{{MN + KN}}\left( {{\lambda _{\cal M}} + {\lambda _{\cal N}}} \right)}},
\end{equation}
\begin{equation} \label{lambda_define}
{\lambda _\cM} \triangleq \sum\limits_{j \in {\cal N}} {\sum\limits_{m = 1}^M {\cu_j^{(m)}} } , \;\; {\lambda _{\cal N}} \triangleq \sum\limits_{j \in {\cal N}} {\sum\limits_{i = 1}^K {u_j^{(i)}} }  \;\; \text{and} \;\; {\lambda _{\cal S}} \triangleq \sum\limits_{j \in {\cal S}} {\sum\limits_{i = 1}^K {u_j^{(i)}} }. 
\end{equation}
It is seen from (\ref{eta_K_M_1_rewritten}) that $\eta _{K,M}^{(1)}$ is determined by the statistics ${\lambda _\cM}$, ${\lambda _{\cal N}} $ and ${\lambda _{\cal S}}$. Thus, if each sensor is able to obtain the values of ${\lambda _\cM}$, ${\lambda _{\cal N}} $ and ${\lambda _{\cal S}}$, then it can produce $\eta _{K,M}^{(1)}$. 
Similarly, by employing (\ref{theta_0_MLE}), (\ref{theta_est}), (\ref{Lambda_G_Simplified}) and (\ref{H_G_Define}), $\eta _{K,M}^{(2)}$ can be written as 
\begin{equation}
\eta _{K,M}^{(2)} = \left( {K{N_{\cal S}} - {\lambda _{\cal S}}} \right)G_{K,M}^{(1)} + {\lambda _{\cal S}}G_{K,M}^{(2)},
\end{equation}
which is fully determined by ${\lambda _\cM}$, ${\lambda _{\cal N}} $ and ${\lambda _{\cal S}}$. Moreover, as shown later, ${\max }_{1 \le k \le K}  \eta _{K,M}^{(4)}$ also depends on the statistics ${\lambda _\cM}$, ${\lambda _{\cal N}} $ and ${\lambda _{\cal S}}$. 
To this end, we consider the distributed computation of  ${\lambda _\cM}$, ${\lambda _{\cal N}} $ and ${\lambda _{\cal S}}$ in the remainder of this subsection.

For the sake of notational simplicity, we first put the quantized data $\{\cu_j^{(m)}\}$, $\{u_j^{(i)}\}$ in three vectors, that is, let $\cbgamma^{(l)}= {[\cgamma _1^{(l)},\cgamma _2^{(l)},...,\cgamma _N^{(l)}]^T}$,  $\bgamma^{(l)} = {[\gamma _1^{(l)},\gamma _2^{(l)},...,\gamma _N^{(l)}]^T}$  and $\tbgamma^{(l)} = {[\tgamma _1^{(l)},\tgamma _2^{(l)},...,\tgamma _N^{(l)}]^T}$ denote the observation vectors at time $l$, where
\begin{equation} \label{cgamma_define}
\cgamma _j^{(l)} \triangleq \left\{ \begin{array}{l}
\cu_j^{(l)}, \quad \text{if} \;\; 1 \le l \le M,\\
0,\qquad \text{if} \;\; l \ge M+1,
\end{array} \right.
\end{equation}
\begin{equation} \label{gamma_define}
\gamma _j^{(l)} \triangleq \left\{ \begin{array}{l}
0, \qquad \text{if} \;\; 1 \le l \le M,\\
u_j^{(l)}, \quad \text{if} \;\; l \ge M+1,
\end{array} \right.
\end{equation}
\begin{equation} \label{tgamma_define}
\tgamma _j^{(l)} \triangleq \left\{ \begin{array}{l}
u_j^{(l)}, \quad \text{if} \;\; l \ge M+1 \text{ and } j \in \cS, \\
0, \qquad \text{otherwise}.\\
\end{array} \right.
\end{equation}

Based on $\cbgamma^{(l)}$, $\bgamma^{(l)}$ and $\tbgamma^{(l)}$,  we can employ running consensus algorithms to obtain   three   local statistics at each sensor which will be utilized to produce local estimates of ${\lambda _\cM}$, ${\lambda _{\cal N}} $ and ${\lambda _{\cal S}}$ later. Specifically, let ${\cbGamma ^{\left( l \right)}}$, ${\bGamma ^{\left( l \right)}}$ and ${\tbGamma ^{\left( l \right)}}$ denote three statistic vectors at time $l$, where the $j$-th element of each statistic vector denotes a local statistic at the $j$-th sensor at time $l$, respectively. We set ${\cbGamma ^{\left( 0 \right)}} = {\bGamma ^{\left( 0 \right)}} = {\tbGamma ^{\left( 0 \right)}} ={\bf 0}$. Within the $l$-th sampling interval, the running-consensus-algorithm-based computation is carried out as follows:
\begin{enumerate}[1)]
	\item Take a new sample at each sensor at time instant $l$, and add it to the statistic vectors from previous time as follows:
	\begin{equation} \label{consensus_sample}
	{\cbGamma ^{\left( {l,0} \right)}} = {\cbGamma ^{\left( {l - 1} \right)}} + {\cbgamma ^{(l)}}, \; {\bGamma ^{\left( {l,0} \right)}} = {\bGamma ^{\left( {l - 1} \right)}} + {\bgamma ^{(l)}},   \text{ and } {\tbGamma ^{\left( {l,0} \right)}} = {\tbGamma ^{\left( {l - 1} \right)}} + {\tbgamma ^{(l)}}.
	\end{equation}
	\item For $n=1,2,...,Q$, every sensor exchanges its local intermediate statistics ${\cbGamma ^{\left( {l,n} \right)}}$, ${\bGamma ^{\left( {l,n} \right)}}$ and ${\tbGamma ^{\left( {l,n} \right)}}$ with its neighboring sensors, and updates the local intermediate statistics as the weighted sum of the available statistics from the neighboring sensors, i.e., for $n=1,2,...,Q$ and $\forall j \in \cN$, 
	\begin{align} \label{message_exchange_check}
	{\left( {{\cbGamma ^{\left( {l,n} \right)}}} \right)_j} & = {\cw_{jj}}{\left( {{\cbGamma ^{\left( {l,n - 1} \right)}}} \right)_j} + \sum\limits_{i \in {{\cal N}_j}} {{\cw_{ij}}{{\left( {{\cbGamma ^{\left( {l,n - 1} \right)}}} \right)}_i}}, \\ \label{message_exchange}
	{\left( {{\bGamma ^{\left( {l,n} \right)}}} \right)_j} & = {w_{jj}}{\left( {{\bGamma ^{\left( {l,n - 1} \right)}}} \right)_j} + \sum\limits_{i \in {{\cal N}_j}} {{w_{ij}}{{\left( {{\bGamma ^{\left( {l,n - 1} \right)}}} \right)}_i}}, \\ \label{message_exchange_tilde}
	 {\left( {{\tbGamma ^{\left( {l,n} \right)}}} \right)_j} & = {\tw_{jj}}{\left( {{\tbGamma ^{\left( {l,n - 1} \right)}}} \right)_j} + \sum\limits_{i \in {{\cal N}_j}} {{\tw_{ij}}{{\left( {{\tbGamma ^{\left( {l,n - 1} \right)}}} \right)}_i}}, 
	\end{align}
	where $\cN_j$ denotes the set of indices of the $j$-th sensor's neighbors and the weight coefficients $\{\cw_{ij}\}$, $\{w_{ij}\}$ and $\{\tw_{ij}\}$ will be specified later. The notation $(\cdot)_j$ in (\ref{message_exchange_check})--(\ref{message_exchange_tilde})  denotes the $j$-th element of the corresponding vector.
	\item Update the local statistics for time $l$ as
	\begin{equation}
	{\cbGamma ^{\left( l \right)}} = {\cbGamma ^{\left( {l,Q} \right)}}, \; {\bGamma ^{\left( l \right)}} = {\bGamma ^{\left( {l,Q} \right)}} \; \text{and} \; {\tbGamma ^{\left( l \right)}} = {\tbGamma ^{\left( {l,Q} \right)}}.
	\end{equation}
	\item Go to Step $1)$ for the next sampling interval $l + 1$.
\end{enumerate}

From (\ref{message_exchange_check})--(\ref{message_exchange_tilde}), we can express each message-passing in a compact matrix form as
\begin{equation} \label{consensus_message_passing}
{\cbGamma ^{\left( {l,n} \right)}} = {\bf{\cW}}{\cbGamma ^{\left( {l,n - 1} \right)}}, \; {\bGamma ^{\left( {l,n} \right)}} = {\bf{W}}{\bGamma ^{\left( {l,n - 1} \right)}} \text{ and } {\tbGamma ^{\left( {l,n} \right)}} = {\bf{\tW}}{\tbGamma ^{\left( {l,n - 1} \right)}}, \text{ for } n=1,2,..,Q,
\end{equation}
where the matrices ${\bf{\cW}} \triangleq \left( {{\cw_{ij}}} \right) \in {{\mathbbm{R}}^{N \times N}}$, ${\bf{W}} \triangleq \left( {{w_{ij}}} \right) \in {{\mathbbm{R}}^{N \times N}}$ and ${\bf{\tW}} \triangleq \left( {{\tw_{ij}}} \right) \in {{\mathbbm{R}}^{N \times N}}$ are formed by $\cw_{ij}$, $w_{ij}$ and $\tw_{ij}$ defined in (\ref{message_exchange_check})--(\ref{message_exchange_tilde}), respectively. Combining (\ref{consensus_sample}) and (\ref{consensus_message_passing}), we know that the statistic vectors evolve over time according to
\begin{equation} \label{consensus_statistic_vector_iteration}
{\cbGamma ^{\left( l \right)}} \! = \! {{\bf{\cW}}^Q}\left( {{\cbGamma ^{\left( {l - 1} \right)}} + {\cbgamma ^{(l)}}} \right), \;  {\bGamma ^{\left( l \right)}} \! =\!  {{\bf{W}}^Q}\left( {{\bGamma ^{\left( {l - 1} \right)}} + {\bgamma ^{(l)}}} \right) \; \text{and} \; {\tbGamma ^{\left( l \right)}} \! = \! {{\bf{\tW}}^Q}\left( {{\tbGamma ^{\left( {l - 1} \right)}} + {\tbgamma ^{(l)}}} \right),  \forall  l \ge 1.
\end{equation}
From (\ref{consensus_statistic_vector_iteration}), the statistic vectors at time $L \ge 1$ can also be equivalently expressed as
\begin{equation} \label{consensus_statistic_vector_time_evolving}
{\cbGamma ^{\left( L \right)}} = \sum\limits_{l = 1}^L {{{\bf{\cW}}^{Q\left( {L - l + 1} \right)}}{\cbgamma ^{(l)}}}, \; {\bGamma ^{\left( L \right)}} = \sum\limits_{l = 1}^L {{{\bf{W}}^{Q\left( {L - l + 1} \right)}}{\bgamma ^{(l)}}} \; \text{and} \; {\tbGamma ^{\left( L \right)}} = \sum\limits_{l = 1}^L {{{\bf{\tW}}^{Q\left( {L - l + 1} \right)}}{\tbgamma ^{(l)}}}.
\end{equation}

Note that (\ref{consensus_statistic_vector_time_evolving}) resembles the consensus algorithm in the fixed-sample-size test where there is no innovation, i.e.,
 $\cbgamma^{(l)}=\bgamma^{(l)}=\tbgamma^{(l)}={\bf 0}$ for $l$ is larger than some $L$. In that case, under certain regularity conditions on ${\bf{\cW}}$, ${\bf{W}}$ and ${\bf{\tW}}$, consensus can be reached in the sense that the elements in ${\cbGamma ^{\left( l \right)}}$, ${\bGamma ^{\left( l \right)}}$ or ${\tbGamma ^{\left( l \right)}}$ converge to the same values as $l \to \infty$, respectively.
In contrast, with the new samples constantly arriving in the running consensus algorithm described in (\ref{consensus_sample})--(\ref{consensus_statistic_vector_time_evolving}), the elements in ${\cbGamma ^{\left( l \right)}}$, ${\bGamma ^{\left( l \right)}}$ or ${\tbGamma ^{\left( l \right)}}$ cannot respectively reach consensus as $l \to \infty$. However, we will show later that even though consensus cannot be reached as $l \to \infty$, we  still can employ ${\cbGamma ^{\left( l \right)}}$, ${\bGamma ^{\left( l \right)}}$ and ${\tbGamma ^{\left( l \right)}}$ to construct local estimators of ${\lambda _\cM}$, ${\lambda _{\cal N}} $ and ${\lambda _{\cal S}}$ at each sensor with bounded estimation   errors,   respectively.

Before proceeding, we first impose the following condition on the weight matrices ${\bf{\cW}}$, ${\bf{W}}$ and ${\bf{\tW}}$.
\begin{condition} \label{Condition_W}
	For the weight matrix $\bW \in \{ {\bf{\cW}},  {\bf{W}}, \bf{\tW} \}$, it satisfies
	\begin{equation}
	{{\bf{1}}^T}{\bf{W}} = {{\bf{1}}^T}, \; {\bf{W1}} = {\bf{1}} \; \text{and} \; 0 < {\sigma _2}\left( {\bf{W}} \right) < 1,
	\end{equation}
	where ${\bf{1}}$ is the all-one vector and ${\sigma _i}( {\bf{W}} )$ denotes the $i$-th largest singular value of $\bf{W}$.
\end{condition}

The condition essentially regulates the elements in the matrix. If we further require that every element in the matrix is non-negative, then Condition \ref{Condition_W} is equivalent to ${\bf W}$ being doubly stochastic.

Based on the statistic vectors ${\cbGamma ^{\left( l \right)}}$, ${\bGamma ^{\left( l \right)}}$ and ${\tbGamma ^{\left( l \right)}}$, for each $j=1,2,...,N$, we consider the following estimators of the statistics ${\lambda _\cM}$, ${\lambda _{\cal N}} $ and ${\lambda _{\cal S}}$, respectively
\begin{equation} \label{hat_lambda_define}
\hat \lambda _{{\cal M},j}^{(K)} \buildrel \Delta \over =  N{\bf{e}}_j^T{\cbGamma ^{\left( {M + K} \right)}}, \;\; \hat \lambda _{{\cal N},j}^{(K)} \buildrel \Delta \over = N{\bf{e}}_j^T{\bGamma ^{\left( {M + K} \right)}} \;\; \text{and} \;\; \hat \lambda _{{\cal S},j}^{(K)} \buildrel \Delta \over = N{\bf{e}}_j^T{\tbGamma ^{\left( {M + K} \right)}},
\end{equation}
where ${{\bf{e}}_j} \buildrel \Delta \over = {[0,...,0,\underbrace 1_{{\text{the }}j\text{-th element}},0,...,0]^T}$. It is worth mentioning that $\hat \lambda _{{\cal M},j}^{(K)}$, $\hat \lambda _{{\cal N},j}^{(K)}$ and $\hat \lambda _{{\cal S},j}^{(K)}$ only employ the local statistics at the $j$-th sensor, and therefore, can be locally obtained at each sensor. Regarding the estimation performance of $\hat \lambda _{{\cal M},j}^{(K)}$, $\hat \lambda _{{\cal N},j}^{(K)}$ and $\hat \lambda _{{\cal S},j}^{(K)}$, we   have   the following lemma.
\begin{lemma} \label{Lemma_lambda_diff_ub}
	For any given $K$, $M$ and $j \in \{1,2,...,N\}$, we have
	\begin{align} \label{lambda_M_diff_ub}
	\left| {\hat \lambda _{{\cal M},j}^{(K)} - {\lambda _{\cM}}} \right| \le {N^{\frac{3}{2}}}\frac{{{{\left[ {{\sigma _2}\left( {\bf{\cW}} \right)} \right]}^Q}}}{{1 - {{\left[ {{\sigma _2}\left( {\bf{\cW}} \right)} \right]}^Q}}}, \\  \label{lambda_N_diff_ub}
	\left| {\hat \lambda _{{\cal N},j}^{(K)} - {\lambda _{\cal N}}} \right| \le {N^{\frac{3}{2}}} \frac{{{{\left[ {{\sigma _2}\left( {\bf{W}} \right)} \right]}^Q}}}{{1 - {{\left[ {{\sigma _2}\left( {\bf{W}} \right)} \right]}^Q}}}, \\  \label{lambda_S_diff_ub}
	\left| {\hat \lambda _{{\cal S},j}^{(K)} - {\lambda _{\cal S}}} \right| \le {N^{\frac{3}{2}}}\frac{{{{\left[ {{\sigma _2}\left( {\bf{\tW}} \right)} \right]}^Q}}}{{1 - {{\left[ {{\sigma _2}( {\bf{\tW}} )} \right]}^Q}}}.
	\end{align}
\end{lemma}
\begin{IEEEproof}
	Refer to Appendix \ref{Proof_Lemma_lambda_diff_ub}.
\end{IEEEproof}

From the definitions of ${\lambda _\cM}$, ${\lambda _{\cal N}}$ and ${\lambda _{\cal S}}$ in (\ref{lambda_define}), we know that the computation of these statistics require an FC to collect all quantized data from all sensors.  However, \emph{Lemma \ref{Lemma_lambda_diff_ub}} illustrates that 
just by employing local communications between neighboring sensors,  each sensor can individually produce some local estimators 
which approximate ${\lambda _{\cM}}$, ${\lambda _{\cN}}$ and ${\lambda _{\cS}}$ very well, since the estimation errors can be bounded from above. Moreover, since $0< {{\sigma _2}( {\bf{\cW}} )}, {{\sigma _2}( {\bf{W}} )}, {{\sigma _2}( {\bf{\tW}} )}<1$, the upper bounds on the estimation errors decrease to $0$ as $Q$ grows to infinity, which implies that these local estimates $\hat \lambda _{{\cal M},j}^{(K)}$, $\hat \lambda _{{\cal N},j}^{(K)}$ and $\hat \lambda _{{\cal S},j}^{(K)}$  can be sufficiently accurate for every $j$ provided that $Q$ is sufficiently large. 

\subsection{Distributed Computation of ${\eta} _{K,M}^{(1)}$ and ${\eta}  _{K,M}^{(2)}$}

By utilizing the local estimators $\hat \lambda _{{\cal M},j}^{(K)}$, $\hat \lambda _{{\cal N},j}^{(K)}$ and $\hat \lambda _{{\cal S},j}^{(K)}$, we define the local estimators of ${\eta} _{K,M}^{(1)}$ and ${\eta}  _{K,M}^{(2)}$ at the $j$-th sensor as
\begin{align} \label{hat_eta_1_define}
\hat \eta _{K,M,j}^{(1)} \buildrel \Delta \over = & \left( {MN - \hat \lambda _{{\cal M},j}^{(K)}} \right)\hat G_{K,M}^{(j,1)} + \hat \lambda _{{\cal M},j}^{(K)}\hat G_{K,M}^{(j,2)}, \\ \label{hat_eta_2_define}
\hat \eta _{K,M,j}^{(2)} \buildrel \Delta \over = & \left( {K{N_{\cal S}} - \hat \lambda _{{\cal S},j}^{(K)}} \right)\hat G_{K,M}^{(j,1)} + \hat \lambda _{{\cal S},j}^{(K)}\hat G_{K,M}^{(j,2)},
\end{align}
respectively, where
\begin{equation} \label{hat_G}
\hat G_{K,M}^{(j,1)} \buildrel \Delta \over = \ln \frac{{1 - \frac{1}{{MN + K{N_{\cal S}}}}\left( {\hat \lambda _{{\cal M},j}^{(K)} + \hat \lambda _{{\cal S},j}^{(K)}} \right)}}{{1 - \frac{1}{{MN + KN}}\left( {\hat \lambda _{{\cal M},j}^{(K)} + \hat \lambda _{{\cal N},j}^{(K)}} \right)}} \; \text{ and } \; \hat G_{K,M}^{(j,2)} \buildrel \Delta \over = \ln \frac{{\frac{1}{{MN + K{N_{\cal S}}}}\left( {\hat \lambda _{{\cal M},j}^{(K)} + \hat \lambda _{{\cal S},j}^{(K)}} \right)}}{{\frac{1}{{MN + KN}}\left( {\hat \lambda _{{\cal M},j}^{(K)} + \hat \lambda _{{\cal N},j}^{(K)}} \right)}}.
\end{equation}

Regarding the asymptotic performance of the estimators $\hat \eta _{K,M,j}^{(1)}$ and $\hat \eta _{K,M,j}^{(2)}$, we provide the following theorem.
\begin{theorem} \label{Theorem_estimate_eta_1_and_2}
	In the asymptotic regime where $M \to \infty$, for any given $K<\infty$ and any sensor $j$, we have
	\begin{align} \notag
	\mathop {\lim }\limits_{M \to \infty } \left| {\hat \eta _{K,M,j}^{(1)} - \eta _{K,M}^{(1)}} \right| \le  & \frac{ {N^2} \left[{2 - q\left( \theta  \right)} \right] }{{q\left( \theta  \right)\left[ {1 - q\left( \theta  \right)} \right]}} \Bigg(  \frac{{2{{\left[ {{\sigma _2}\left( {\bf{\cW}} \right)} \right]}^Q}}}{{1 - {{\left[ {{\sigma _2}\left( {\bf{\cW}} \right)} \right]}^Q}}}  + \frac{{{{\left[ {{\sigma _2}\left( {\bf{\tW}} \right)} \right]}^Q}}}{{1 - {{\left[ {{\sigma _2}\left( {\bf{\tW}} \right)} \right]}^Q}}} \\  \label{theorem_eta_diff_1}
	&  \qquad  + \frac{{{{\left[ {{\sigma _2}\left( {\bf{W}} \right)} \right]}^Q}}}{{1 - {{\left[ {{\sigma _2}\left( {\bf{W}} \right)} \right]}^Q}}}\Bigg) \quad \text{ a.s.} 
	\end{align}
	and
	\begin{equation} \label{theorem_eta_diff_2}
	\mathop {\lim }\limits_{M \to \infty } \left| {\hat \eta _{K,M,j}^{(2)} - \eta _{K,M}^{(2)}} \right| = 0,	\quad \text{ a.s.}.
	\end{equation}
\end{theorem}
\begin{IEEEproof}
From (\ref{eta_K_M_1_rewritten}) and (\ref{hat_eta_1_define}), we can obtain
\begin{align} \notag
\left| {\hat \eta _{K,M,j}^{(1)} - \eta _{K,M}^{(1)}} \right| 
& = \left| {\left( {MN - \hat \lambda _{{\cal M},j}^{(K)}} \right)\hat G_{K,M}^{(j,1)} + \hat \lambda _{{\cal M},j}^{(K)}\hat G_{K,M}^{(j,2)} - \left( {MN - {\lambda _{\cal M}}} \right)G_{K,M}^{(1)} - {\lambda _{\cal M}}G_{K,M}^{(2)}} \right|\\ \label{eta_diff_ub_1}
& \le \left| {\left( {MN - \hat \lambda _{{\cal M},j}^{(K)}} \right)\hat G_{K,M}^{(j,1)} - \left( {MN - {\lambda _{\cal M}}} \right)G_{K,M}^{(1)}} \right| \! + \! \left| {\hat \lambda _{{\cal M},j}^{(K)}\hat G_{K,M}^{(j,2)} - {\lambda _{\cal M}}G_{K,M}^{(2)}} \right|.
\end{align}
Regarding the first term in (\ref{eta_diff_ub_1}), we have
\begin{align} \notag
& \left| {\left( {MN - \hat \lambda _{{\cal M},j}^{(K)}} \right)\hat G_{K,M}^{(j,1)} - \left( {MN - {\lambda _{\cal M}}} \right)G_{K,M}^{(1)}} \right|\\ \notag
& = \left| {\left( {MN - \hat \lambda _{{\cal M},j}^{(K)}} \right)\hat G_{K,M}^{(j,1)} - \left( {MN - \hat \lambda _{{\cal M},j}^{(K)}} \right)G_{K,M}^{(1)} + \left( {MN - \hat \lambda _{{\cal M},j}^{(K)}} \right)G_{K,M}^{(1)} - \left( {MN - {\lambda _{\cal M}}} \right)G_{K,M}^{(1)}} \right|\\ \notag
& \le \left| {\left( {MN - \hat \lambda _{{\cal M},j}^{(K)}} \right)\hat G_{K,M}^{(j,1)} - \left( {MN - \hat \lambda _{{\cal M},j}^{(K)}} \right)G_{K,M}^{(1)}} \right| + \left| {\left( {MN - \hat \lambda _{{\cal M},j}^{(K)}} \right)G_{K,M}^{(1)} - \left( {MN - {\lambda _{\cal M}}} \right)G_{K,M}^{(1)}} \right|\\ \label{eta_diff_1st_term_ub_1}
& = \left| {MN - \hat \lambda _{{\cal M},j}^{(K)}} \right|\left| {\hat G_{K,M}^{(j,1)} - G_{K,M}^{(1)}} \right| + \left| {\hat \lambda _{{\cal M},j}^{(K)} - {\lambda _{\cal M}}} \right|\left| {G_{K,M}^{(1)}} \right|.
\end{align}
Similarly, the second term in (\ref{eta_diff_ub_1}) can be bounded from above as per
\begin{align}  \label{eta_diff_2nd_term_ub_1}
\left| {\hat \lambda _{{\cal M},j}^{(K)}\hat G_{K,M}^{(j,2)} - {\lambda _{\cal M}}G_{K,M}^{(2)}} \right| \le \left| {\hat \lambda _{{\cal M},j}^{(K)}} \right|\left| {\hat G_{K,M}^{(j,2)} - G_{K,M}^{(2)}} \right| + \left| {\hat \lambda _{{\cal M},j}^{(K)} - {\lambda _{\cal M}}} \right|\left| {G_{K,M}^{(2)}} \right|.
\end{align}
From (\ref{consensus_statistic_vector_time_evolving}) and the definition of ${\hat \lambda _{{\cal M},j}^{(K)}}$ in (\ref{hat_lambda_define}), we know that 
\begin{align} \notag
\left| {\hat \lambda _{{\cal M},j}^{(K)}} \right| & = \left| {N{\bf{e}}_j^T{\cbGamma ^{\left( {M + K} \right)}}} \right|\\ \notag
& = \left| {N{\bf{e}}_j^T\sum\limits_{l = 1}^{M + K} {{{\bf{\cW}}^{Q\left( {M + K - l + 1} \right)}}{\cbgamma ^{(l)}}} } \right|\\ \label{hat_lambda_abs_ub_temp_1}
& \le N{\left\| {{{\bf{e}}_j}} \right\|_2}{\left\| {{{\bf{\cW}}^{QK}}\sum\limits_{l = 1}^M {{{\bf{\cW}}^{Q\left( {M - l + 1} \right)}}{\cbgamma ^{(l)}}} } \right\|_2}\\ \label{hat_lambda_abs_ub_temp_2}
& \le N\sum\limits_{l = 1}^M {{{\left\| {{{\bf{\cW}}^{Q\left( {M - l + 1} \right)}}{\cbgamma ^{(l)}}} \right\|}_2}} \\ \label{hat_lambda_abs_ub}
& \le M{N^{\frac{3}{2}}},
\end{align}
which implies
\begin{equation} \label{MN_lambda_abs_ub}
\left| {MN - \hat \lambda _{{\cal M},j}^{(K)}} \right| \le MN + \left| {\hat \lambda _{{\cal M},j}^{(K)}} \right| \le 2M{N^{\frac{3}{2}}}.
\end{equation}
The inequality in (\ref{hat_lambda_abs_ub_temp_1}) is due to the Cauchy-Schwarz inequality and (\ref{cgamma_define}). The inequalities in (\ref{hat_lambda_abs_ub_temp_2}) and (\ref{hat_lambda_abs_ub}) are from the fact that $|{\gamma_j ^{(l)}}|\le 1$ and Condition \ref{Condition_W} implies ${\sigma _1}( {\bf{\cW}} ) = 1$.

By employing (\ref{eta_diff_ub_1}), (\ref{eta_diff_1st_term_ub_1}), (\ref{eta_diff_2nd_term_ub_1}), (\ref{hat_lambda_abs_ub}), (\ref{MN_lambda_abs_ub}) and Lemma \ref{Lemma_lambda_diff_ub}, we can obtain
\begin{align} \notag
\left| {\hat \eta _{K,M,j}^{(1)} - \eta _{K,M}^{(1)}} \right| \le &  2M{N^{\frac{3}{2}}}\left| {\hat G_{K,M}^{(j,1)} - G_{K,M}^{(1)}} \right| + M{N^{\frac{3}{2}}}\left| {\hat G_{K,M}^{(j,2)} - G_{K,M}^{(2)}} \right| \\ \label{eta_diff_ub_2}
& \qquad  + {N^{\frac{3}{2}}}\frac{{{{\left[ {{\sigma _2}\left( {\bf{\cW}} \right)} \right]}^Q}}}{{1 - {{\left[ {{\sigma _2}\left( {\bf{\cW}} \right)} \right]}^Q}}}\left( {\left| {G_{K,M}^{(1)}} \right| + \left| {G_{K,M}^{(2)}} \right|} \right),
\end{align}
which implies
\begin{align} \notag
\mathop {\lim }\limits_{M \to \infty } \left| {\hat \eta _{K,M,j}^{(1)} - \eta _{K,M}^{(1)}} \right| &  \le {N^{\frac{3}{2}}}\mathop {\lim }\limits_{M \to \infty } M\left( {2\left| {\hat G_{K,M}^{(j,1)} - G_{K,M}^{(1)}} \right| + \left| {\hat G_{K,M}^{(j,2)} - G_{K,M}^{(2)}} \right|} \right) + \\ \label{limit_eta_diff_ub_1}
& \qquad  {N^{\frac{3}{2}}}\frac{{{{\left[ {{\sigma _2}\left( {\bf{\cW}} \right)} \right]}^Q}}}{{1 - {{\left[ {{\sigma _2}\left( {\bf{\cW}} \right)} \right]}^Q}}}\left( {\mathop {\lim }\limits_{M \to \infty } \left| {G_{K,M}^{(1)}} \right| + \mathop {\lim }\limits_{M \to \infty } \left| {G_{K,M}^{(2)}} \right|} \right).
\end{align}

We first consider the term ${| {\hat G_{K,M}^{(j,1)} - G_{K,M}^{(1)}} |}$ in (\ref{limit_eta_diff_ub_1}). It is clear that
\begin{align} \notag
\left| {\hat G_{K,M}^{(j,1)} - G_{K,M}^{(1)}} \right| & = \left| {\ln \frac{{1 - \frac{1}{{MN + K{N_{\cal S}}}}\left( {\hat \lambda _{{\cal M},j}^{(K)} + \hat \lambda _{{\cal S},j}^{(K)}} \right)}}{{1 - \frac{1}{{MN + KN}}\left( {\hat \lambda _{{\cal M},j}^{(K)} + \hat \lambda _{{\cal N},j}^{(K)}} \right)}} - \ln \frac{{1 - \frac{1}{{MN + K{N_{\cal S}}}}\left( {{\lambda _{\cal M}} + {\lambda _{\cal S}}} \right)}}{{1 - \frac{1}{{MN + KN}}\left( {{\lambda _{\cal M}} + {\lambda _{\cal N}}} \right)}}} \right|\\ \notag
& = \left| {\ln \frac{{1 - \frac{1}{{MN + K{N_{\cal S}}}}\left( {\hat \lambda _{{\cal M},j}^{(K)} + \hat \lambda _{{\cal S},j}^{(K)}} \right)}}{{1 - \frac{1}{{MN + K{N_{\cal S}}}}\left( {{\lambda _{\cal M}} + {\lambda _{\cal S}}} \right)}} - \ln \frac{{1 - \frac{1}{{MN + KN}}\left( {\hat \lambda _{{\cal M},j}^{(K)} + \hat \lambda _{{\cal N},j}^{(K)}} \right)}}{{1 - \frac{1}{{MN + KN}}\left( {{\lambda _{\cal M}} + {\lambda _{\cal N}}} \right)}}} \right|\\ \label{G_1_diff_ub}
& \le \left| {\ln \frac{{1 - \frac{1}{{MN + K{N_{\cal S}}}}\left( {\hat \lambda _{{\cal M},j}^{(K)} + \hat \lambda _{{\cal S},j}^{(K)}} \right)}}{{1 - \frac{1}{{MN + K{N_{\cal S}}}}\left( {{\lambda _{\cal M}} + {\lambda _{\cal S}}} \right)}}} \right| + \left| {\ln \frac{{1 - \frac{1}{{MN + KN}}\left( {\hat \lambda _{{\cal M},j}^{(K)} + \hat \lambda _{{\cal N},j}^{(K)}} \right)}}{{1 - \frac{1}{{MN + KN}}\left( {{\lambda _{\cal M}} + {\lambda _{\cal N}}} \right)}}} \right|.
\end{align}

Noting that $1 - 1/x \le \ln x \le x - 1$
and by employing Lemma \ref{Lemma_lambda_diff_ub}, we can obtain
\begin{align} \notag
\left| {\ln \frac{{1 - \frac{1}{{MN + K{N_{\cal S}}}}\left( {\hat \lambda _{{\cal M},j}^{(K)} + \hat \lambda _{{\cal S},j}^{(K)}} \right)}}{{1 - \frac{1}{{MN + K{N_{\cal S}}}}\left( {{\lambda _{\cal M}} + {\lambda _{\cal S}}} \right)}}} \right| & \le   \frac{{A_{K,M}^{(1,1)}}}{{MN + K{N_{\cal S}}}}\left| {\hat \lambda _{{\cal M},j}^{(K)} + \hat \lambda _{{\cal S},j}^{(K)} - {\lambda _{\cal M}} - {\lambda _{\cal S}}} \right| \\ \notag
& \le \frac{{A_{K,M}^{(1,1)}}}{{MN + K{N_{\cal S}}}}\left( {\left| {\hat \lambda _{{\cal M},j}^{(K)} - {\lambda _{\cal M}}} \right| + \left| {\hat \lambda _{{\cal S},j}^{(K)} - {\lambda _{\cal S}}} \right|} \right)\\  \label{G_1_1st_term_diff_ub}
& \le \frac{{A_{K,M}^{(1,1)}{N^{\frac{3}{2}}}}}{{MN + K{N_{\cal S}}}}\left( {\frac{{{{\left[ {{\sigma _2}\left( {\bf{\cW}} \right)} \right]}^Q}}}{{1 - {{\left[ {{\sigma _2}\left( {\bf{\cW}} \right)} \right]}^Q}}} + \frac{{{{\left[ {{\sigma _2}\left( {\bf{\tW}} \right)} \right]}^Q}}}{{1 - {{\left[ {{\sigma _2}\left( {\bf{\tW}} \right)} \right]}^Q}}}} \right),
\end{align}
where the quantity $A_{K,M}^{(1,1)}$ is defined as
\begin{equation} \label{A_1_1_define}
A_{K,M}^{(1,1)} \buildrel \Delta \over = \max \left\{ {\frac{1}{{\left| {1 - \frac{1}{{MN + K{N_{\cal S}}}}\left( {{\lambda _{\cal M}} + {\lambda _{\cal S}}} \right)} \right|}},\frac{1}{{\left| {1 - \frac{1}{{MN + K{N_{\cal S}}}}\left( {\hat \lambda _{{\cal M},j}^{(K)} + \hat \lambda _{{\cal S},j}^{(K)}} \right)} \right|}}} \right\}.
\end{equation}
Similarly, the second term in (\ref{G_1_diff_ub}) can be bounded from above as per
\begin{equation}  \label{G_1_2nd_term_diff_ub}
\left| {\ln \frac{{1 - \frac{1}{{MN + KN}}\left( {\hat \lambda _{{\cal M},j}^{(K)} + \hat \lambda _{{\cal N},j}^{(K)}} \right)}}{{1 - \frac{1}{{MN + KN}}\left( {{\lambda _{\cal M}} + {\lambda _{\cal N}}} \right)}}} \right| \le \frac{{A_{K,M}^{(1,2)}{N^{\frac{3}{2}}}}}{{MN + KN}}\left( {\frac{{{{\left[ {{\sigma _2}\left( {\bf{\cW}} \right)} \right]}^Q}}}{{1 - {{\left[ {{\sigma _2}\left( {\bf{\cW}} \right)} \right]}^Q}}} + \frac{{{{\left[ {{\sigma _2}\left( {\bf{W}} \right)} \right]}^Q}}}{{1 - {{\left[ {{\sigma _2}\left( {\bf{W}} \right)} \right]}^Q}}}} \right),
\end{equation}
where $A_{K,M}^{(1,2)}$ is defined as
\begin{equation} \label{A_1_2_define}
A_{K,M}^{(1,2)} \buildrel \Delta \over = \max \left\{ {\frac{1}{{\left| {1 - \frac{1}{{MN + KN}}\left( {{\lambda _{\cal M}} + {\lambda _{\cal N}}} \right)} \right|}},\frac{1}{{\left| {1 - \frac{1}{{MN + KN}}\left( {\hat \lambda _{{\cal M},j}^{(K)} + \hat \lambda _{{\cal N},j}^{(K)}} \right)} \right|}}} \right\},
\end{equation}
which yields
\begin{align} \notag
\left| {\hat G_{K,M}^{(j,1)} - G_{K,M}^{(1)}} \right|  \le & \frac{{A_{K,M}^{(1,1)}{N^{\frac{3}{2}}}}}{{MN + K{N_{\cal S}}}}\left( {\frac{{{{\left[ {{\sigma _2}\left( {\bf{\cW}} \right)} \right]}^Q}}}{{1 - {{\left[ {{\sigma _2}\left( {\bf{\cW}} \right)} \right]}^Q}}} + \frac{{{{\left[ {{\sigma _2}\left( {\bf{\tW}} \right)} \right]}^Q}}}{{1 - {{\left[ {{\sigma _2}\left( {\bf{\tW}} \right)} \right]}^Q}}}} \right) \\ \label{G_1_diff_ub_final}
& \quad  + \frac{{A_{K,M}^{(1,2)}{N^{\frac{3}{2}}}}}{{MN + KN}}\left( {\frac{{{{\left[ {{\sigma _2}\left( {\bf{\cW}} \right)} \right]}^Q}}}{{1 - {{\left[ {{\sigma _2}\left( {\bf{\cW}} \right)} \right]}^Q}}} + \frac{{{{\left[ {{\sigma _2}\left( {\bf{W}} \right)} \right]}^Q}}}{{1 - {{\left[ {{\sigma _2}\left( {\bf{W}} \right)} \right]}^Q}}}} \right)
\end{align}
by employing (\ref{G_1_diff_ub}) and (\ref{G_1_1st_term_diff_ub}).

Note that
\begin{align} \notag
& \left| {1 - \frac{1}{{MN + K{N_{\cal S}}}}\left( {\hat \lambda _{{\cal M},j}^{(K)} + \hat \lambda _{{\cal S},j}^{(K)}} \right)} \right| \\ \notag
&  \ge \left| {1 - \frac{1}{{MN + K{N_{\cal S}}}}\left( {{\lambda _{\cal M}} + {\lambda _{\cal S}}} \right)} \right| - \left| {\frac{1}{{MN + K{N_{\cal S}}}}\left( {\hat \lambda _{{\cal M},j}^{(K)} - {\lambda _{\cal M}} + \hat \lambda _{{\cal S},j}^{(K)} - {\lambda _{\cal S}}} \right)} \right|\\ \label{temp_1}
& \ge \left| {1 - \frac{1}{{MN + K{N_{\cal S}}}}\left( {{\lambda _{\cal M}} + {\lambda _{\cal S}}} \right)} \right| - \frac{{{N^{\frac{3}{2}}}}}{{MN + K{N_{\cal S}}}}\left( {\frac{{{{\left[ {{\sigma _2}\left( {\bf{\cW}} \right)} \right]}^Q}}}{{1 - {{\left[ {{\sigma _2}\left( {\bf{\cW}} \right)} \right]}^Q}}} + \frac{{{{\left[ {{\sigma _2}\left( {\bf{\tW}} \right)} \right]}^Q}}}{{1 - {{\left[ {{\sigma _2}\left( {\bf{\tW}} \right)} \right]}^Q}}}} \right),
\end{align}
which implies
\begin{align} \notag
\mathop {\lim }\limits_{M \to \infty } \left| {1 - \frac{1}{{MN + K{N_{\cal S}}}}\left( {\hat \lambda _{{\cal M},j}^{(K)} + \hat \lambda _{{\cal S},j}^{(K)}} \right)} \right| & \ge \mathop {\lim }\limits_{M \to \infty } \left| {1 - \frac{1}{{MN + K{N_{\cal S}}}}\left( {{\lambda _{\cal M}} + {\lambda _{\cal S}}} \right)} \right| \\ \label{temp_2}
& = q\left( \theta  \right) \quad \text{ a.s.}
\end{align}
by employing (\ref{q_define}), (\ref{lambda_define}) and the fact that the second term in (\ref{temp_1}) decreases to $0$ as $M \to \infty$. As a result, from (\ref{A_1_1_define}) and (\ref{temp_2}), we can obtain
\begin{align} \notag
\mathop {\lim }\limits_{M \to \infty } A_{K,M}^{(1,1)} & =  \max \left\{ {\mathop {\lim }\limits_{M \to \infty }\frac{1}{{\left| {1 - \frac{1}{{MN + K{N_{\cal S}}}}\left( {{\lambda _{\cal M}} + {\lambda _{\cal S}}} \right)} \right|}},\mathop {\lim }\limits_{M \to \infty }\frac{1}{{\left| {1 - \frac{1}{{MN + K{N_{\cal S}}}}\left( {\hat \lambda _{{\cal M},j}^{(K)} + \hat \lambda _{{\cal S},j}^{(K)}} \right)} \right|}}} \right\}\\ \label{limit_A_1_1}
& = \frac{1}{{q\left( \theta  \right)}} \quad \text{ a.s.}
\end{align}
Similarly, we can obtain
\begin{equation}  \label{limit_A_1_2}
\mathop {\lim }\limits_{M \to \infty } A_{K,M}^{(1,2)} = \frac{1}{{q\left( \theta  \right)}} \quad \text{ a.s.}.
\end{equation}
Furthermore, by employing (\ref{G_1_diff_ub_final}), (\ref{limit_A_1_1}) and (\ref{limit_A_1_2}), we have
\begin{align} \notag
& \mathop {\lim }\limits_{M \to \infty } M\left| {\hat G_{K,M}^{(j,1)} - G_{K,M}^{(1)}} \right| \\ \notag
& \le \mathop {\lim }\limits_{M \to \infty } \frac{{A_{K,M}^{(1,1)}M{N^{\frac{3}{2}}}}}{{MN + K{N_{\cal S}}}}\left( {\frac{{{{\left[ {{\sigma _2}\left( {\bf{\cW}} \right)} \right]}^Q}}}{{1 - {{\left[ {{\sigma _2}\left( {\bf{\cW}} \right)} \right]}^Q}}} + \frac{{{{\left[ {{\sigma _2}\left( {\bf{\tW}} \right)} \right]}^Q}}}{{1 - {{\left[ {{\sigma _2}\left( {\bf{\tW}} \right)} \right]}^Q}}}} \right) \\ \notag
& \qquad + \mathop {\lim }\limits_{M \to \infty } \frac{{A_{K,M}^{(1,2)}M{N^{\frac{3}{2}}}}}{{MN + KN}}\left( {\frac{{{{\left[ {{\sigma _2}\left( {\bf{\cW}} \right)} \right]}^Q}}}{{1 - {{\left[ {{\sigma _2}\left( {\bf{\cW}} \right)} \right]}^Q}}} + \frac{{{{\left[ {{\sigma _2}\left( {\bf{W}} \right)} \right]}^Q}}}{{1 - {{\left[ {{\sigma _2}\left( {\bf{W}} \right)} \right]}^Q}}}} \right)\\  \label{limit_G_diff_1}
& = \frac{{N^{\frac{1}{2}}}}{{q\left( \theta  \right)}}\left( {\frac{{2{{\left[ {{\sigma _2}\left( {\bf{\cW}} \right)} \right]}^Q}}}{{1 - {{\left[ {{\sigma _2}\left( {\bf{\cW}} \right)} \right]}^Q}}} + \frac{{{{\left[ {{\sigma _2}\left( {\bf{\tW}} \right)} \right]}^Q}}}{{1 - {{\left[ {{\sigma _2}\left( {\bf{\tW}} \right)} \right]}^Q}}} + \frac{{{{\left[ {{\sigma _2}\left( {\bf{W}} \right)} \right]}^Q}}}{{1 - {{\left[ {{\sigma _2}\left( {\bf{W}} \right)} \right]}^Q}}}} \right) \quad \text{ a.s.}.
\end{align}

By following a similar approach as above, we can obtain
\begin{align} \notag
& \mathop {\lim }\limits_{M \to \infty } M\left| {\hat G_{K,M}^{(j,2)} - G_{K,M}^{(2)}} \right|\\ \label{limit_G_diff_2}
& \le \frac{{N^{\frac{1}{2}}}}{1-{q\left( \theta  \right)}}\left( {\frac{{2{{\left[ {{\sigma _2}\left( {\bf{\cW}} \right)} \right]}^Q}}}{{1 - {{\left[ {{\sigma _2}\left( {\bf{\cW}} \right)} \right]}^Q}}} + \frac{{{{\left[ {{\sigma _2}\left( {\bf{\tW}} \right)} \right]}^Q}}}{{1 - {{\left[ {{\sigma _2}\left( {\bf{\tW}} \right)} \right]}^Q}}} + \frac{{{{\left[ {{\sigma _2}\left( {\bf{W}} \right)} \right]}^Q}}}{{1 - {{\left[ {{\sigma _2}\left( {\bf{W}} \right)} \right]}^Q}}}} \right) \quad \text{ a.s.}.
\end{align}

  Moreover,    we also have
\begin{equation} \label{limit_G_1}
\mathop {\lim }\limits_{M \to \infty } \left| {G_{K,M}^{(1)}} \right| = \left| {\mathop {\lim }\limits_{M \to \infty } \ln \frac{{1 - \frac{1}{{MN + K{N_{\cal S}}}}\left( {{\lambda _{\cal M}} + {\lambda _{\cal S}}} \right)}}{{1 - \frac{1}{{MN + KN}}\left( {{\lambda _{\cal M}} + {\lambda _{\cal N}}} \right)}}} \right| = \left| {\ln \frac{{F\left( {\tau  - \theta } \right)}}{{F\left( {\tau  - \theta } \right)}}} \right| = 0 \quad \text{ a.s.},
\end{equation}
\begin{equation} \label{limit_G_2}
\mathop {\lim }\limits_{M \to \infty } \left| {G_{K,M}^{(2)}} \right| = \left| {\mathop {\lim }\limits_{M \to \infty } \ln \frac{{\frac{1}{{MN + K{N_{\cal S}}}}\left( {{\lambda _{\cal M}} + {\lambda _{\cal S}}} \right)}}{{\frac{1}{{MN + KN}}\left( {{\lambda _{\cal M}} + {\lambda _{\cal N}}} \right)}}} \right| = \left| {\ln \frac{{1 - F\left( {\tau  - \theta } \right)}}{{1 - F\left( {\tau  - \theta } \right)}}} \right| = 0 \quad \text{ a.s.}.
\end{equation}
Therefore, from (\ref{limit_eta_diff_ub_1}), (\ref{limit_G_diff_1}), (\ref{limit_G_diff_2}), (\ref{limit_G_1}) and (\ref{limit_G_2}), we can obtain
\begin{align} \notag
& \mathop {\lim }\limits_{M \to \infty } \left| {\hat \eta _{K,M,j}^{(1)} - \eta _{K,M}^{(1)}} \right| \\ \notag
& \le {N^{\frac{3}{2}}}\mathop {\lim }\limits_{M \to \infty } M\left( {2\left| {\hat G_{K,M}^{(j,1)} - G_{K,M}^{(1)}} \right|  + \left| {\hat G_{K,M}^{(j,2)} - G_{K,M}^{(2)}} \right|} \right) \\ \notag
& \qquad \qquad  + {N^{\frac{3}{2}}}\frac{{{{\left[ {{\sigma _2}\left( {\bf{\cW}} \right)} \right]}^Q}}}{{1 - {{\left[ {{\sigma _2}\left( {\bf{\cW}} \right)} \right]}^Q}}}\left( {\mathop {\lim }\limits_{M \to \infty } \left| {G_{K,M}^{(1)}} \right| + \mathop {\lim }\limits_{M \to \infty } \left| {G_{K,M}^{(2)}} \right|} \right)\\ \notag
& = {N^{\frac{3}{2}}}\Bigg\{ \frac{2}{{q\left( \theta  \right)}}{N^{\frac{1}{2}}}\Bigg( {\frac{{2{{\left[ {{\sigma _2}\left( {\bf{\cW}} \right)} \right]}^Q}}}{{1 - {{\left[ {{\sigma _2}\left( {\bf{\cW}} \right)} \right]}^Q}}} + \frac{{{{\left[ {{\sigma _2}\left( {\bf{\tW}} \right)} \right]}^Q}}}{{1 - {{\left[ {{\sigma _2}\left( {\bf{\tW}} \right)} \right]}^Q}}} + \frac{{{{\left[ {{\sigma _2}\left( {\bf{W}} \right)} \right]}^Q}}}{{1 - {{\left[ {{\sigma _2}\left( {\bf{W}} \right)} \right]}^Q}}}} \Bigg) \\  \notag
& \qquad \qquad + \frac{1}{{1 - q\left( \theta  \right)}}{N^{\frac{1}{2}}}\Bigg( {\frac{{2{{\left[ {{\sigma _2}\left( {\bf{\cW}} \right)} \right]}^Q}}}{{1 - {{\left[ {{\sigma _2}\left( {\bf{\cW}} \right)} \right]}^Q}}} + \frac{{{{\left[ {{\sigma _2}\left( {\bf{\tW}} \right)} \right]}^Q}}}{{1 - {{\left[ {{\sigma _2}\left( {\bf{\tW}} \right)} \right]}^Q}}} + \frac{{{{\left[ {{\sigma _2}\left( {\bf{W}} \right)} \right]}^Q}}}{{1 - {{\left[ {{\sigma _2}\left( {\bf{W}} \right)} \right]}^Q}}}} \Bigg) \Bigg\}\\
& = \frac{ {N^2} \left[{2 - q\left( \theta  \right)} \right] }{{q\left( \theta  \right)\left[ {1 - q\left( \theta  \right)} \right]}}\left( {\frac{{2{{\left[ {{\sigma _2}\left( {\bf{\cW}} \right)} \right]}^Q}}}{{1 - {{\left[ {{\sigma _2}\left( {\bf{\cW}} \right)} \right]}^Q}}} + \frac{{{{\left[ {{\sigma _2}\left( {\bf{\tW}} \right)} \right]}^Q}}}{{1 - {{\left[ {{\sigma _2}\left( {\bf{\tW}} \right)} \right]}^Q}}} + \frac{{{{\left[ {{\sigma _2}\left( {\bf{W}} \right)} \right]}^Q}}}{{1 - {{\left[ {{\sigma _2}\left( {\bf{W}} \right)} \right]}^Q}}}} \right) \quad \text{ a.s.},
\end{align}
which completes the proof for (\ref{theorem_eta_diff_1}).

Next, we prove (\ref{theorem_eta_diff_2}).
Similar to (\ref{eta_diff_ub_1}), (\ref{eta_diff_1st_term_ub_1}) and (\ref{eta_diff_2nd_term_ub_1}), we can obtain
\begin{align} \notag
\left| {\hat \eta _{K,M,j}^{(2)} - \eta _{K,M}^{(2)}} \right| &  = \left| {\left( {K{N_{\cal S}} - \hat \lambda _{{\cal S},j}^{(K)}} \right)\hat G_{K,M}^{(j,1)} + \hat \lambda _{{\cal S},j}^{(K)}\hat G_{K,M}^{(j,2)} - \left( {K{N_{\cal S}} - {\lambda _{\cal S}}} \right)G_{K,M}^{(1)} - {\lambda _{\cal S}}G_{K,M}^{(2)}} \right| \\ \notag
& \le \left| {K{N_{\cal S}} - \hat \lambda _{{\cal S},j}^{(K)}} \right|\left| {\hat G_{K,M}^{(j,1)} - G_{K,M}^{(1)}} \right| + \left| {\hat \lambda _{{\cal S},j}^{(K)} - {\lambda _{\cal S}}} \right|\left| {G_{K,M}^{(1)}} \right| \\ \label{eta_2_diff_ub_1}
& \qquad + \left| {\hat \lambda _{{\cal S},j}^{(K)}} \right|\left| {\hat G_{K,M}^{(j,2)} - G_{K,M}^{(2)}} \right| + \left| {\hat \lambda _{{\cal S},j}^{(K)} - {\lambda _{\cal S}}} \right|\left| {G_{K,M}^{(2)}} \right|
\end{align}
by employing the definitions of ${\eta _{K,M}^{(2)}}$ and ${\hat \eta _{K,M,j}^{(2)}}$ in (\ref{H_A_Define}) and (\ref{hat_eta_2_define}), respectively.

Note that since ${\sigma _1}( {\bf{\tW}} ) = 1$, $|{\widetilde{\gamma}}_j^{(l)}|\le 1$, and  ${\widetilde{\gamma}}_j^{(l)}=0$ for $l\le M$, 
\begin{align} \notag
\left| {\hat \lambda _{{\cal S},j}^{(K)}} \right| & = \left| {N{\bf{e}}_j^T{\tbGamma ^{\left( {M + K} \right)}}} \right|\\ \notag
& = \left| {N{\bf{e}}_j^T\sum\limits_{l = 1}^{M + K} {{{\bf{\tW}}^{Q\left( {M + K - l + 1} \right)}}{\tbgamma ^{(l)}}} } \right|\\ \notag
& \le N{\left\| {{{\bf{e}}_j}} \right\|_2}{\left\| {\sum\limits_{l = M + 1}^{M + K} {{{\bf{\tW}}^{Q\left( {M + K - l + 1} \right)}}{\tbgamma ^{(l)}}} } \right\|_2}\\ \notag
& \le N\sum\limits_{l = M + 1}^{M + K} {{{\left\| {{{\bf{\tW}}^{Q\left( {M + K - l + 1} \right)}}{\tbgamma ^{(l)}}} \right\|}_2}} \\
& \le K{N^{\frac{3}{2}}},
\end{align}
which yields
\begin{align} \notag
\left| {\hat \eta _{K,M,j}^{(2)} - \eta _{K,M}^{(2)}} \right| & \le 2K{N^{\frac{3}{2}}}\left| {\hat G_{K,M}^{(j,1)} - G_{K,M}^{(1)}} \right| + K{N^{\frac{3}{2}}}\left| {\hat G_{K,M}^{(j,2)} - G_{K,M}^{(2)}} \right| \\ \label{eta_2_diff_ub_2}
& \qquad + {N^{\frac{3}{2}}}\frac{{{{\left[ {{\sigma _2}\left( {\bf{W}} \right)} \right]}^Q}}}{{1 - {{\left[ {{\sigma _2}\left( {\bf{W}} \right)} \right]}^Q}}}\left( {\left| {G_{K,M}^{(1)}} \right| + \left| {G_{K,M}^{(2)}} \right|} \right)
\end{align}
by employing (\ref{eta_2_diff_ub_1}) and Lemma \ref{Lemma_lambda_diff_ub}. As a result, from (\ref{limit_G_diff_1}), (\ref{limit_G_diff_2}), (\ref{limit_G_1}) and (\ref{limit_G_2}), we know
\begin{equation}
\mathop {\lim }\limits_{M \to \infty } \left| {\hat \eta _{K,M,j}^{(2)} - \eta _{K,M}^{(2)}} \right| = 0 \quad \text{ a.s.}
\end{equation}
which completes the proof.
\end{IEEEproof}

Notably, the estimation   performances of $\hat{\eta} _{K,M,j}^{(1)}$ and $\hat{\eta}  _{K,M,j}^{(2)}$ are   different in the asymptotic regime where $M \to \infty$. Theorem \ref{Theorem_estimate_eta_1_and_2} demonstrates that as $M \to \infty$, the estimation error of $\hat{\eta}  _{K,M,j}^{(2)}$ decreases to $0$ with probability one, while the estimation error of $\hat{\eta} _{K,M,j}^{(1)}$ only can be bounded above with probability one. This is mainly because, as shown by (\ref{hat_eta_1_define}) and (\ref{hat_eta_2_define}), $\hat{\eta} _{K,M,j}^{(1)}$ is scalable with respect to $M$ while $\hat{\eta}  _{K,M,j}^{(2)}$ is not. 
However, the estimation error of $\hat{\eta} _{K,M,j}^{(1)}$ still can be controlled just by increasing the number of message-passing   rounds   $Q$ in each sampling interval, i.e., as $Q \to \infty$, the upper bounds in (\ref{theorem_eta_diff_1}) and (\ref{theorem_eta_diff_2}) decrease to $0$.



\subsection{Distributed Computation of $ {\max }_{1 \le k \le K}  \eta _{K,M}^{(4)}$}

At last, we consider the distributed computation of the term $ {\max }_{1 \le k \le K} \eta _{K,M}^{(4)}$ in (\ref{H_A_Define}).
From (\ref{H_A_Define}), we know $ {\max }_{1 \le k \le K} \eta _{K,M}^{(4)} = {\max }_{1 \le k \le K} \sum_{i = k}^K {\sum_{j \in {\cal A}} {\phi _{ji}^{\left( 4 \right)}} } $ which implies that all the attacked sensors are stricken at the same time, i.e., they share the same estimate of the attack time. This underlying assumption inhibits the distributed computation of $ {\max }_{1 \le k \le K} \sum_{i = k}^K {\sum_{j \in {\cal A}} {\phi _{ji}^{\left( 4 \right)}} } $, since estimating the attack time (the maximization operation over $k$) requires all sensor data up to time $K$ can be   processed   at a single place.
To this end,   we relax the constraint of simultaneous attack and assume that the attack times at different sensors  are different.   In consequence, we modify the term  $ {\max }_{1 \le k \le K} \sum_{i = k}^K {\sum_{j \in {\cal A}} {\phi _{ji}^{\left( 4 \right)}} } $ by allowing each sensor to search its own attack time, that is
\begin{equation} \label{relaxation}
\mathop {\max }\limits_{1 \le k \le K} \eta _{K,M}^{(4)} = \mathop {\max }\limits_{1 \le k \le K} \sum\limits_{i = k}^K {\sum\limits_{j \in {\cal A}} {\phi _{ji}^{\left( 4 \right)}} }  \Rightarrow \mathop {\max }\limits_{1 \le k \le K} \eta _{K,M}^{(4)} \approx \sum\limits_{j \in {\cal A}} {\underbrace {\mathop {\max }\limits_{1 \le k_j \le K} \sum\limits_{i = k_j}^K {\phi _{ji}^{\left( 4 \right)}} }_{\triangleq \psi _j^{(K)}}}.
\end{equation}
It is worth mentioning that in the case where adversaries cannot perfectly synchronously launch attacks at sensors  (e.g., attack times at different sensors can be distinct), searching the attack time at each individual sensor is expected to be more reasonable and effective. Moreover, this idea of allowing each sensor to search its own attack time is widely employed in the quickest detection in sensor networks, see \cite{mei2010efficient, li2015quickest, liu2017distributed} for instance.


The quantity $\psi _j^{(K)}$ in (\ref{relaxation}) can be recursively expressed as
\begin{equation} \label{psi_recursive}
\psi _j^{(K)} = \mathop {\max }\limits_{1 \le {k_j} \le K} \sum\limits_{i = {k_j}}^K {\phi _{ji}^{\left( 4 \right)}} 
=  \max \bigg\{ {\mathop {\max }\limits_{1 \le {k_j} \le K - 1} \sum\limits_{i = {k_j}}^K {\phi _{ji}^{\left( 4 \right)}} ,\phi _{jK}^{\left( 4 \right)}} \bigg\} 
= \max \left\{ {\psi _j^{(K - 1)},0} \right\} + \phi _{jK}^{\left( 4 \right)}
\end{equation}
with $\psi _j^{(0)}=0$.
Moreover, from (\ref{Lambda_G_Simplified}), ${\phi _{ji}^{\left( 4 \right)}}$ can be rewritten as
\begin{align} \notag
\phi _{ji}^{\left( 4 \right)} & = \ln \frac{{{\bbP_1}\left( {u_j^{(i)}\left| {{{\hat \theta }^{(a)}},{{\hat \mu }_j}} \right.} \right)}}{{{\bbP_0}\left( {u_j^{(i)}\left| {\hat \theta _{{\rm{MLE}}}^{(u)}} \right.} \right)}}\\  \label{psi_ji_4_simplified}
& = \left( {1 - u_j^{(i)}} \right)\ln \frac{{\zeta _j^{(K)}}}{{1 - \frac{1}{{\left( {M + K} \right)N}}\left( {{\lambda _{\cal M}} + {\lambda _{\cal N}}} \right)}} + u_j^{(i)}\ln \frac{{1 - \zeta _j^{(K)}}}{{\frac{1}{{\left( {M + K} \right)N}}\left( {{\lambda _{\cal M}} + {\lambda _{\cal N}}} \right)}},
\end{align}
where the statistic ${\zeta _j^{(K)}}$ is defined as
\begin{align} \notag
\zeta _j^{(K)} \triangleq & \bone\left\{ {{{\tilde \mu }_j} \ge b} \right\}\left( {1 - \frac{1}{{K - {k_j} + 1}}\sum\limits_{i = {k_j}}^K {u_j^{(i)}} } \right) \\ \label{zeta_jK_define}
& \qquad + \bone\left\{ {{{\tilde \mu }_j} < b} \right\}F\left( {{F^{ - 1}}\left( {1  - \frac{1}{{{MN + KN_{\cS}}}}\left( {{\lambda _{\cal M}} + {\lambda _{\cal S}}} \right)} \right) - b} \right),
\end{align}
and ${{\tilde \mu }_j}$ is given by (\ref{tilde_mu_approx}) except that $k$ is replaced by $k_j$.

In order to compute $\phi _{ji}^{\left( 4 \right)}$ locally at the $j$-th sensor, motivated by Lemma \ref{Lemma_lambda_diff_ub}, we employ the estimators ${\hat \lambda _{{\cal M},j}^{(i)}}$  and ${\hat \lambda _{{\cal S},j}^{(i)}}$ as respective substitutes for  ${\lambda _{\cal M}}$ and ${\lambda _{\cal N}}$ in the definition of ${{\tilde \mu }_j}$,  (\ref{psi_ji_4_simplified}) and (\ref{zeta_jK_define}). In addition, noticing that the term ${\frac{1}{{K - {k_j} + 1}}\sum_{i = {k_j}}^K {u_j^{(i)}} }$ in (\ref{zeta_jK_define}) only utilizes more recent data and ignores the  earlier data,
we can employ 
\begin{equation} \label{hat_lambda_A}
\hat \lambda _{{\cal A},j}^{(K)} \buildrel \Delta \over = \frac{1}{{\sum\limits_{l = 1}^K {{\alpha ^{K - l}}} }}\sum\limits_{l = 1}^K {{\alpha ^{K - l}}u_j^{(l)}}  = \frac{{1 - \alpha }}{{1 - {\alpha ^K}}}\sum\limits_{l = 1}^K {{\alpha ^{K - l}}u_j^{(l)}} \;\; \text{with  } \alpha<1
\end{equation} 
to approximate ${\frac{1}{{K - {k_j} + 1}}\sum_{i = {k_j}}^K {u_j^{(i)}} }$ which is from the widespread intuitive idea of exponential weighting of observations which uses higher weights on the recent observations and lower weights on past ones \cite{basseville1993detection, tartakovsky2014sequential}. 
As such, ${\phi _{ji}^{\left( 4 \right)}}$ can be approximated by $\hat \phi _{ji}^{\left( 4 \right)}$ which can be expressed as
\begin{align} \label{hat_phi_ji_4}
\hat \phi _{ji}^{\left( 4 \right)} = \left( {1 - u_j^{(i)}} \right)\ln \frac{{\hat \zeta _j^{(i)}}}{{1 - \frac{1}{{\left( {M + i} \right)N}}\left( {\hat \lambda _{{\cal M},j}^{(i)} + \hat \lambda _{{\cal N},j}^{(i)}} \right)}} + u_j^{(i)}\ln \frac{{1 - \hat \zeta _j^{(i)}}}{{\frac{1}{{\left( {M + i} \right)N}}\left( {\hat \lambda _{{\cal M},j}^{(i)} + \hat \lambda _{{\cal N},j}^{(i)}} \right)}},
\end{align}
where ${\hat \zeta _j^{(i)}}$ is defined as
\begin{align} \label{hat_zeta_ji}
\hat \zeta _j^{(i)} \! \buildrel \Delta \over = \! \bone\left\{ {\tilde \mu _j^{(i)} \ge b} \right\}\left( {1 - \hat \lambda _{{\cal A},j}^{(i)}} \right) \! + \! \bone\left\{ {\tilde \mu _j^{(i)} \! < \! b} \right\}F\left( {{F^{ - 1}}\left( {1 \! - \! \frac{1}{{ {MN + iN_{\cS}} }}\left( {\hat \lambda _{{\cal M},j}^{(i)} + \hat \lambda _{{\cal S},j}^{(i)}} \right)} \right) \! - \!  b} \right)
\end{align}
and ${\tilde \mu _j^{(i)}}$ is given by
\begin{equation} \label{tilde_mu_j_i}
\tilde \mu _j^{(i)} \buildrel \Delta \over = {F^{ - 1}}\left( {1 - \frac{1}{{{MN + iN_{\cS}}}}\left( {\hat \lambda _{{\cal M},j}^{(i)} + \hat \lambda _{{\cal S},j}^{(i)}} \right)} \right) - {F^{ - 1}}\left( {1 - \hat \lambda _{{\cal A},j}^{(i)}} \right).
\end{equation}


It is seen from (\ref{hat_phi_ji_4}), (\ref{hat_zeta_ji}) and (\ref{tilde_mu_j_i}) that $\hat \phi _{ji}^{\left( 4 \right)}$ can be calculated locally by using local estimators. Furthermore, from (\ref{hat_lambda_A}), the local statistic  $\hat \lambda _{{\cal A},j}^{(i)}$ can be recursively written as
\begin{align}
\hat \lambda _{{\cal A},j}^{(i)} = \frac{{1 - \alpha }}{{1 - {\alpha ^i}}}\left( {\alpha \sum\limits_{l = 1}^{i - 1} {{\alpha ^{K - l - 1}}u_j^{(l)}}  + u_j^{(i)}} \right) = \frac{{\alpha  - {\alpha ^i}}}{{1 - {\alpha ^i}}}\hat \lambda _{{\cal A},j}^{(i - 1)} + \frac{{1 - \alpha }}{{1 - {\alpha ^i}}}u_j^{(i)},
\end{align}
which can further facilitate the distributed computation based on running consensus algorithms.

By employing (\ref{psi_recursive}) and (\ref{hat_phi_ji_4}), the corresponding estimate $\hat \psi _j^{(K)}$ of $\psi _j^{(K)}$ can be expressed as
\begin{equation}
\hat \psi _j^{(K)} = \mathop {\max }\limits_{1 \le {k_j} \le K} \sum\limits_{i = {k_j}}^K {\hat \phi _{ji}^{\left( 4 \right)}}  = \max \left\{ {\hat \psi _j^{(K - 1)},0} \right\} + \hat \phi _{jK}^{\left( 4 \right)} \;\; \text{ with } \;\; \hat \psi _j^{(0)}=0,
\end{equation}
which yields that $ {\max }_{1 \le k \le K} \eta _{K,M}^{(4)}$ can be approximated by
\begin{equation} \label{bar_eta}
\bar \eta _{K,M}^{(3)} \buildrel \Delta \over = \sum\limits_{j \in {\cal A}} {\mathop {\max }\limits_{1 \le {k_j} \le K} \sum\limits_{i = {k_j}}^K {\hat \phi _{ji}^{\left( 4 \right)}} }  = \sum\limits_{j \in {\cal A}} {\hat \psi _j^{(K)}}  = \sum\limits_{j \in {\cal A}} {\sum\limits_{l = 1}^{M+K} {\xi _j^{(l)}} }
\end{equation}
where the $j$-th element $\xi _j^{(l)}$ of the vector ${{{\bxi }}^{(i)}} = {[\xi _1^{(i)},\xi _2^{(i)},...,\xi _N^{(i)}]^T}$ is defined as
\begin{equation} \label{xi_j_i_define}
\xi _j^{(i)} \buildrel \Delta \over = \left\{ \begin{array}{l}
\hat \psi _j^{(i - M)} - \hat \psi _j^{(i - M - 1)}, \; i \ge M + 1 \text{ and } j \in {\cal A}, \\
0, \quad  \text{otherwise},
\end{array} \right.
\end{equation}
It is seen from (\ref{bar_eta}) that $\bar \eta _{K,M}^{(3)}$ is a sum of local statistics over sensors and over time. Thus, similar to the estimator $\hat \lambda _{{\cal M},j}^{(i)}$, by using (\ref{xi_j_i_define}), we also can employ the running consensus algorithm to compute $\bar \eta _{K,M}^{(3)}$ at each sensor in a distributed way 
\begin{equation} \label{hat_eat_3_define}
\hat \eta _{K,M,j}^{(3)} \buildrel \Delta \over = {\bf{e}}_j^T\sum\limits_{l = 1}^{M+K} {{{{\bf{\bar W}}}^{Q\left( {M+ K - l + 1} \right)}}{{\boldsymbol{\xi }}^{(l)}}} = {\bf{e}}_j^T\sum\limits_{l = 1}^{K} {{{{\bf{\bar W}}}^{Q\left( {K - l + 1} \right)}}{{\boldsymbol{\xi }}^{(l+M)}}}.
\end{equation}

\subsection{Distributed Approximate Generalized CUSUM}

Based on (\ref{hat_eta_1_define}), (\ref{hat_eta_2_define}) and (\ref{hat_eat_3_define}), we define the local test statistic at the $j$-th sensor as follows
\begin{equation} \label{H_D_Define}
H_{K,M,j}^{\left( D \right)} \buildrel \Delta \over = \hat \eta _{K,M,j}^{(1)} + \hat \eta _{K,M,j}^{(2)} + \hat \eta _{K,M,j}^{(3)}.
\end{equation}
Consequently, the distributed approximate generalized CUSUM (DAG-CUSUM) test at the $j$-th sensor can be written as
\begin{equation} \label{DAGCUSUM}
T_j^{(D)} = \min \left\{ {K:H_{K,M,j}^{\left( D \right)} \ge {h}} \right\}.
\end{equation}

Comparing $H_{K,M}^{\left( A \right)}$ in (\ref{H_A_Define}) with $H_{K,M,j}^{\left( D \right)}$ in (\ref{H_D_Define}), we know
\begin{equation}
\left| {H_{K,M,j}^{\left( D \right)} - H_{K,M}^{\left( A \right)}} \right| \le \left| {\hat \eta _{K,M,j}^{(1)} - \eta _{K,M}^{(1)}} \right| + \left| {\hat \eta _{K,M,j}^{(2)} - \eta _{K,M}^{(2)}} \right| + \left| {\hat \eta _{K,M,j}^{(3)} - \mathop {\max }\limits_{1 \le k \le K} \eta _{K,M}^{(4)}} \right|
\end{equation}
which yields
\begin{align} \notag
\mathop {\lim }\limits_{M \to \infty } \left| {H_{K,M,j}^{\left( D \right)} - H_{K,M}^{\left( A \right)}} \right| \le & \frac{ {N^2} \left[{2 - q\left( \theta  \right)} \right] }{{q\left( \theta  \right)\left[ {1 - q\left( \theta  \right)} \right]}} \Bigg(  \frac{{2{{\left[ {{\sigma _2}\left( {\bf{\cW}} \right)} \right]}^Q}}}{{1 - {{\left[ {{\sigma _2}\left( {\bf{\cW}} \right)} \right]}^Q}}}  + \frac{{{{\left[ {{\sigma _2}\left( {\bf{\tW}} \right)} \right]}^Q}}}{{1 - {{\left[ {{\sigma _2}\left( {\bf{\tW}} \right)} \right]}^Q}}} \\   \label{H_D_H_A_Diff}
&  \quad  + \frac{{{{\left[ {{\sigma _2}\left( {\bf{W}} \right)} \right]}^Q}}}{{1 - {{\left[ {{\sigma _2}\left( {\bf{W}} \right)} \right]}^Q}}}\Bigg)   + \mathop {\lim }\limits_{M \to \infty } \left| {\hat \eta _{K,M,j}^{(3)} - \mathop {\max }\limits_{1 \le k \le K} \eta _{K,M}^{(4)}} \right| \quad \text{ a.s.}
\end{align}
by employing Theorem \ref{Theorem_estimate_eta_1_and_2}. The first term on the right-hand side of (\ref{H_D_H_A_Diff}) is a constant. Thus, in the case where $h$ is sufficiently large, the relative error induced by the first term can be neglected, since when alarm is triggered, ${H_{K,M,j}^{\left( D \right)}, \; H_{K,M}^{\left( A \right)}} \ge h$. In addition, the first term can be reduced to arbitrarily small just by increasing the number of message-passings $Q$ in each sampling interval. To this end, 
in the asymptotic regime where $M \to \infty$, the difference between $H_{K,M}^{\left( A \right)}$ and $H_{K,M,j}^{\left( D \right)}$ is mainly determined by the term ${\lim }_{M \to \infty } | {\hat \eta _{K,M,j}^{(3)} -  {\max }_{1 \le k \le K} \eta _{K,M}^{(4)}} |$ which might induce possible performance loss.

Due to the relaxation employed in (\ref{relaxation}), it seems impossible to analytically characterize ${\lim }_{M \to \infty } | {\hat \eta _{K,M,j}^{(3)}} -  {\max }_{1 \le k \le K} \eta _{K,M}^{(4)} |$. As shown in the numerical results in the next section, the performance of the DAG-CUSUM in (\ref{DAGCUSUM}) is comparable to that of GCUSUM in (\ref{GCUSUM_define}).
It is worth mentioning that the proposed DAG-CUSUM features some merits when compared to the GCUSUM. To be specific, it is seen from (\ref{GCUSUM_define}) that for each $K$, the computation of the test statistic of the GCUSUM requires an exhaustive search over $k$. Thus, if the stopping time of the GCUSUM is equal to $K$, the computational complexity of GCUSUM is $\cO(K^2)$. In contrast, since the test statistic of DAG-CUSUM can be written in a recursive form, the computational complexity of the DAG-CUSUM is only $\cO(K)$, which implies that a significant computational complexity reduction can be achieved when employing the DAG-CUSUM, especially in the case where $h$ is large. Moreover, as mentioned before, the GCUSUM requires centralized processing, while the DAG-CUSUM can be implemented in a distributed way. Therefore, the long-distance communication overheads can be spared when employing the DAG-CUSUM. 
We summarize the procedure for computing the DAG-CUSUM at one sensor in Algorithm \ref{Algorithm_DAGCUSUM}.
\begin{algorithm}[htb]
	\caption{Distributed computation of the DAG-CUSUM at the $j$-th sensor}
	\begin{algorithmic}[1]
		\STATE \textbf{Input}:  $\{\cu_j^{(m)}\}$, $\{u_j^{(i)}\}$, ${\bf{\cW}}$,  ${\bf{W}}$, $\bf{\tW}$, $\bf{\bar W}$, $Q$, $\alpha$ and $h$;
		\STATE  \textbf{Output}: $T_j^{(D)}$;
		\STATE $\cbGamma^{(0)} \leftarrow {\bf 0}$;
		\FOR{$m=1$ to $M$}
		\STATE Evaluate $\cbgamma^{(m)}$ by using (\ref{cgamma_define});
		\STATE Using $\cbgamma^{(m)}$ to evaluate $\cbGamma^{(m)}$ by employing the running consensus algorithm with the weight matrix ${\bf{\cW}}$ as in (\ref{consensus_statistic_vector_time_evolving});
		\ENDFOR
		\STATE $\hat \lambda _{{\cal M},j}^{(0)} \leftarrow N{\bf{e}}_j^T{\cbGamma ^{\left( M \right)}}$, $\hat \lambda _{{\cal S},j}^{(0)} \leftarrow 0 $ and  $\hat \lambda _{{\cal N},j}^{(0)} \leftarrow 0 $;
		\STATE Using $\hat \lambda _{{\cal M},j}^{(0)}$, $\hat \lambda _{{\cal S},j}^{(0)}$ and $\hat \lambda _{{\cal N},j}^{(0)}$ to evaluate ${\hat \eta _{0,M,j}^{(1)}}$ as in (\ref{hat_eta_1_define});
		\STATE $H_{0,M,j}^{\left( D \right)} \leftarrow \hat \eta _{0,M,j}^{(1)}$ and $T_j^{(D)} \leftarrow 0$;
		\STATE $\bGamma^{(M)} \leftarrow {\bf 0}$, $\tbGamma^{(M)} \leftarrow {\bf 0}$, $\hat \lambda _{{\cal A},j}^{(0)} \leftarrow 0$ and $K  \leftarrow 0$;
		\WHILE {$H_{K,M,j}^{\left( D \right)} < h$}
		\STATE $K \leftarrow K +1$;
		\STATE Evaluate $\cbgamma^{(M+K)}$, $\bgamma^{(M+K)}$ and $\tbgamma^{(M+K)}$ by using (\ref{cgamma_define}), (\ref{gamma_define}) and (\ref{tgamma_define}), respectively;
		\STATE Using $\cbgamma^{(M+K)}$, $\bgamma^{(M+K)}$ and $\tbgamma^{(M+K)}$ to evaluate $\cbGamma^{(M+K)}$, $\bGamma^{(M+K)}$ and $\tbGamma^{(M+K)}$ by employing the running consensus algorithm with ${\bf{\cW}}$,  ${\bf{W}}$ and $\bf{\tW}$ as in (\ref{consensus_statistic_vector_time_evolving});
		\STATE  $\hat \lambda _{{\cal M},j}^{(K)} \leftarrow N{\bf{e}}_j^T{\cbGamma ^{\left( M+K \right)}}$, $\hat \lambda _{{\cal S},j}^{(K)} \leftarrow N{\bf{e}}_j^T{\bGamma ^{\left( M+K \right)}}$ and $\hat \lambda _{{\cal N},j}^{(K)} \leftarrow N{\bf{e}}_j^T{\tbGamma ^{\left( M+K \right)}}$;
		\STATE $\hat \lambda _{{\cal A},j}^{(K)}  \leftarrow \frac{{\alpha  - {\alpha ^i}}}{{1 - {\alpha ^i}}}\hat \lambda _{{\cal A},j}^{(K - 1)} + \frac{{1 - \alpha }}{{1 - {\alpha ^i}}}u_j^{(K)}$;
		\STATE Using $\hat \lambda _{{\cal M},j}^{(K)}$, $\hat \lambda _{{\cal S},j}^{(K)}$ and $\hat \lambda _{{\cal N},j}^{(K)}$ to evaluate ${\hat \eta _{K,M,j}^{(1)}}$,  ${\hat \eta _{K,M,j}^{(2)}}$ as in (\ref{hat_eta_1_define}) and (\ref{hat_eta_2_define}), respectively;
		\STATE Using $\hat \lambda _{{\cal A},j}^{(K)}$,  $\hat \lambda _{{\cal M},j}^{(K)}$, $\hat \lambda _{{\cal S},j}^{(K)}$ and $\hat \lambda _{{\cal N},j}^{(K)}$ to evaluate $\bxi^{(K)}$ as in (\ref{xi_j_i_define});
		\STATE Using $\bxi^{(K)}$ to evaluate ${\hat \eta _{K,M,j}^{(3)}}$ by employing the running consensus algorithm with ${\bf{\bar W}}$ as in (\ref{hat_eat_3_define});
		\STATE $H_{K,M,j}^{\left( D \right)} \leftarrow \hat \eta _{K,M,j}^{(1)} + \hat \eta _{K,M,j}^{(2)} + \hat \eta _{K,M,j}^{(3)}$;
		\STATE $T_j^{(D)} \leftarrow K$;
		\ENDWHILE
	\end{algorithmic}
	\label{Algorithm_DAGCUSUM}
\end{algorithm}

\section{Numerical Results}
\label{Section_Numerical_Results}

In this section, we examine the performance of the DAG-CUSUM in (\ref{DAGCUSUM}) and compare it to that of GCUSUM in (\ref{GCUSUM_define}). 

\subsection{Choice of $\bf \cW$, $\bW$ and $\bf \tW$}

First, we specify the weight matrices $\bf \cW$, $\bW$ and $\bf \tW$ in (\ref{consensus_statistic_vector_iteration}). There are multiple
methods to choose W such that Condition 1 can be satisfied, one of which is assigning equal
weights to the data from neighbors \cite{xiao2004fast}. Here, we employ a similar idea as that in \cite{xiao2004fast} and choose
${\bf \cW} =\bW = {\bf \tW}$ 
In specific, the weight matrix $\bW$ admits
\begin{equation} \label{W_Numerical}
{\bf{W}} = {\bf{I}} -\frac{2\left( {{\bf{D}} - {\bf{A}}} \right)}{ \sigma _1^2\left( {{\bf{D}} - {\bf{A}}} \right) + \sigma _{N - 1}^2\left( {{\bf{D}} - {\bf{A}}} \right)} ,
\end{equation}
where ${\bf{A}}$ is the adjacent matrix, whose entries $a_{ij} = 1$ if and only if ${i, j} \in \cE$ and the matrix ${\bf D} \triangleq {\text{diag}}\{|\cN_1|, |\cN_2|, . . . , |\cN_K|\}$ is called degree matrix. Their difference is called Laplacian matrix which is positive semidefinite. It can be shown that $\bW$ in (\ref{W_Numerical}) satisfies Condition \ref{Condition_W} \cite{xiao2004fast}.

\subsection{Simulation Setup}

\begin{figure}[htb]
	\centerline{
		\includegraphics[width=0.46\textwidth]{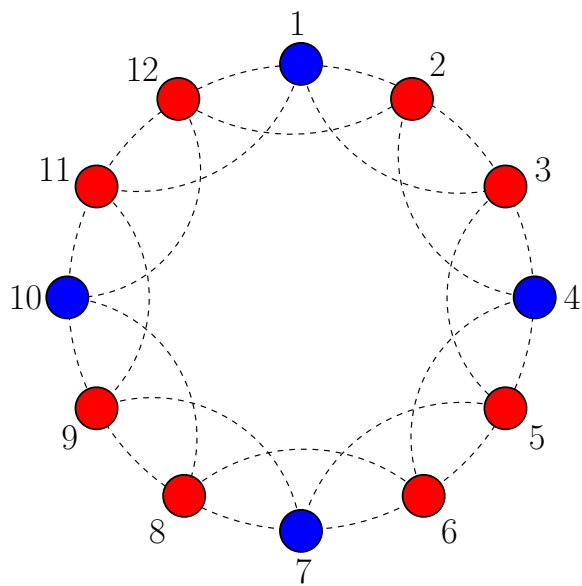}
	}
	\caption{The topology of the sensor network.}		
	\label{Fig_Sensor_Network}	
\end{figure}

Consider a sensor network consisting of $12$ sensors which is depicted in Fig. \ref{Fig_Sensor_Network}. The blue circles represent secure sensors and the red ones are insecure. The edges of the graph are represented by the dash lines. The weight matrix in (\ref{W_Numerical}) for the graph in Fig. \ref{Fig_Sensor_Network} has $\sigma_2(W)=0.6511$. 

In the simulation, the unknown parameter $\theta = 1$, the threshold of the quantizers is $\tau=1$, and the injected data $\mu_j = 0.2$ for all $j$. The noise $n_j^{(k)}$ in (\ref{x_j_k_model}) obeys Gaussian distribution with zero mean and unit variance for all $j$ and $k$.  Moreover, the lower bound $b$ on the injected data in (\ref{mu_constraint}) is $0.18$,  the attack time $t_a$ is set to $10$. In addition, for the DAG-CUSUM, $Q=10$, the parameter $\alpha = 0.979$ and   $M=5 \times 10^3$.  

\vspace{3mm}

\begin{figure}[htb]
	\centerline{
		\includegraphics[width=0.86\textwidth]{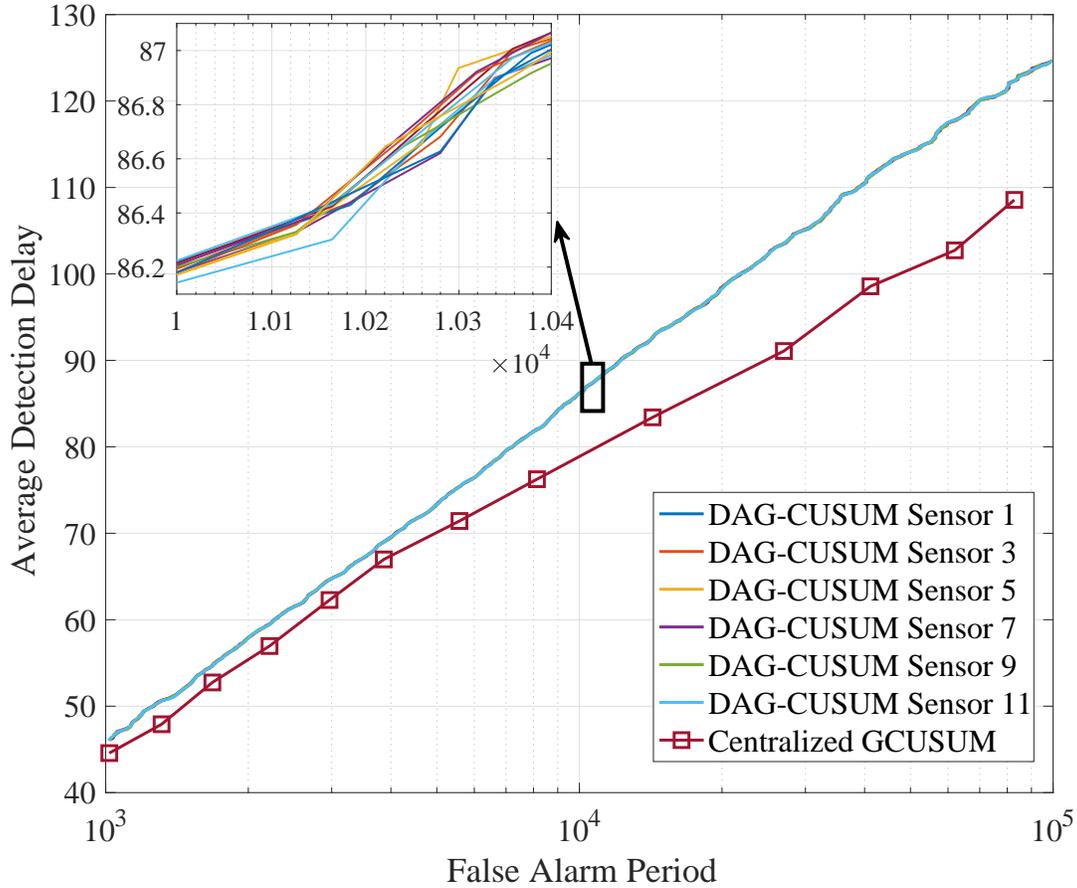}
	}
	\caption{Performance comparison between the centralized GCUSUM and the DAG-CUSUM at different sensor.}			
	\label{Fig_Performance_Comparison}	
\end{figure}

In Fig. \ref{Fig_Performance_Comparison}, we compare the proposed DAG-CUSUM in (\ref{DAGCUSUM}) with the centralized GCUSUM in (\ref{GCUSUM_define}) in terms of the average detection delay as the false alarm period increases. The number of Monte Carlo runs is $2000$. Fig. \ref{Fig_Performance_Comparison} illustrates that when the DAG-CUSUM is employed, for any given false alarm period, the average detection delay of each sensor is very close to that of other sensors, which implies that the sensors, no matter attacked or unattacked, can detect the occurrence of attacks almost simultaneously. It is worth mentioning that it takes way more time to obtain the curve for the centralized GCUSUM than the curves for the DAG-CUSUM.   Specifically, the running time for obtaining the curve for the centralized GCUSUM is about $242$ hours; while the running time for obtaining the curves for the DAG-CUSUM is just $742$ seconds on the same computer.   
It is seen from Fig. \ref{Fig_Performance_Comparison} that by employing the DAG-CUSUM, we reduce the computational complexity and eliminate the global communication overheads  at the price of some performance loss, since the performance of the centralized GCUSUM outperforms that of every sensor when the DAG-CUSUM is employed. This performance loss is possibly mainly induced by the relaxation in (\ref{relaxation}). 


\section{Conclusions}
\label{Section_Conclusions}

  We have considered the generalized CUSUM algorithm for change detection in  the presence of unknown parameters in both the pre-change and post-change models, in the context of quickest attack detection in a distributed parameter estimation system with one-bit quantized measurements. 
First, a sufficient condition is provided under which the expected false alarm period of the GCUSUM test can be guaranteed to be larger than any given value.  
Since the centralized GCUSUM incurs significant communication overhead, and moreover, has a prohibitively high computational complexity, we therefore focus on developing the distributed implementation of the GCUSUM. An alternative test statistic is first proposed  as a substitute for that of GCUSUM which can be shown to be asymptotically equivalent. Then based on the proposed alternative  test statistic and running consensus algorithms, we have proposed the DAG-CUSUM which significantly  reduce the computational complexity and requires only local communications between neighboring sensors. 
The numerical results show that the proposed DAG-CUSUM algorithm can provide a performance that is comparable to the centralized  GCUSUM.

\appendices

\section{Proof of Lemma \ref{Lemma_lambda_diff_ub}} \label{Proof_Lemma_lambda_diff_ub}

We will just prove (\ref{lambda_M_diff_ub}), and the   proofs for (\ref{lambda_N_diff_ub}) and (\ref{lambda_S_diff_ub}) are similar.  

By employing (\ref{cgamma_define}), (\ref{lambda_define}) and (\ref{hat_lambda_define}), we can obtain
\begin{equation} \label{lemma_temp_1}
\left| {\hat \lambda _{{\cal M},j}^{(K)} - {\lambda _{\cM}}} \right| = \left| {N{\bf{e}}_j^T{\cbGamma ^{\left( {M + K} \right)}} - \sum\limits_{i \in {\cal N}} {\sum\limits_{m = 1}^M {\cu_i^{(m)}} } } \right| = \left| {N{\bf{e}}_j^T\left( {{\cbGamma ^{\left( {M + K} \right)}} - \sum\limits_{m = 1}^M {{\bf{J}}{\cbgamma ^{(l)}}} } \right)} \right|,
\end{equation}	
where the matrix $\bf J$ is defined as
\begin{equation} \label{J_define}
{\bf J} \triangleq \frac{{\bf 1} {\bf 1}^T}{N}.
\end{equation}
From the definition of $\bf J$ in (\ref{J_define}) and Condition \ref{Condition_W}, we have
\begin{equation} \label{W_J_property}
{{\bf{\cW}}^n} - {\bf{J}} = {\left( {{\bf{\cW}} - {\bf{J}}} \right)^n}, \; \forall n=1,2,...,
\end{equation}
which can be proved by induction as follows:1) For $n=1$, it is clear that (\ref{W_J_property}) is true; 2) assume that ${{\bf{\cW}}^n} - {\bf{J}} = {\left( {{\bf{\cW}} - {\bf{J}}} \right)^n}$ for some $n$, then
\begin{align} \notag
{\left( {{\bf{\cW}} - {\bf{J}}} \right)^{n + 1}} & = \left( {{\bf{\cW}} - {\bf{J}}} \right){\left( {{\bf{\cW}} - {\bf{J}}} \right)^n}\\  \notag
& = \left( {{\bf{\cW}} - {\bf{J}}} \right)\left( {{{\bf{\cW}}^n} - {\bf{J}}} \right)\\  \notag
& = {{\bf{\cW}}^{n + 1}} - {\bf{\cW J}} - {\bf{J}}{{\bf{\cW}}^n} + {{\bf{J}}^2}\\ \label{lemma_W_J_temp_1}
& = {{\bf{\cW}}^{n + 1}} - {\bf{J}},
\end{align}
where (\ref{lemma_W_J_temp_1}) is due to the fact that ${{\bf{J}}^2} = {\bf{J}}$ and ${\bf{\cW J}} = {\bf{J}}{{\bf{\cW}}^n} = {\bf{J}}$ by employing (\ref{J_define}) and Condition \ref{Condition_W}. As a result, by employing (\ref{consensus_statistic_vector_time_evolving}),  (\ref{lemma_temp_1}) can be simplified as
\begin{align} \notag
\left| {\hat \lambda _{{\cal M},j}^{(K)} - {\lambda _{\cM}}} \right| & = \left| {N{\bf{e}}_j^T\left( {{\cbGamma ^{\left( {M + K} \right)}} - \sum\limits_{m = 1}^M {{\bf{J}}{\cbgamma ^{(l)}}} } \right)} \right|\\ \notag
& = \left| {N{\bf{e}}_j^T\sum\limits_{l = 1}^{M + K} {\left( {{{\bf{\cW}}^{Q\left( {M + K - l + 1} \right)}} - {\bf{J}}} \right){\cbgamma ^{(l)}}} } \right|\\ \notag
& = \left| {N{\bf{e}}_j^T\sum\limits_{l = 1}^{M + K} {{{\left( {{\bf{\cW}} - {\bf{J}}} \right)}^{Q\left( {M + K - l + 1} \right)}}{\cbgamma ^{(l)}}} } \right|\\ \label{lemma_W_J_temp_2}
& \le N{\left\| {{{\bf{e}}_j}} \right\|_2}{\left\| {\sum\limits_{l = 1}^{M + K} {{{\left( {{\bf{\cW}} - {\bf{J}}} \right)}^{Q\left( {M + K - l + 1} \right)}}{\cbgamma ^{(l)}}} } \right\|_2} \\ \label{lemma_W_J_temp_3}
&  \le N\sum\limits_{l = 1}^{M + K} {{{\left\| {{{\left( {{\bf{\cW}} - {\bf{J}}} \right)}^{Q\left( {M + K - l + 1} \right)}}{\cbgamma ^{(l)}}} \right\|}_2}},
\end{align}
where (\ref{lemma_W_J_temp_2}) is from the Cauchy-Schwarz inequality.

Noting that ${\left\| {{\bf{Ax}}} \right\|_2} \le {\sigma _1}\left( {\bf{A}} \right){\left\| {\bf{x}} \right\|_2}$ for any matrix $\bf W$ and vector $\bf x$, and Condition \ref{Condition_W} implies 
\begin{equation}
{\sigma _1}\left( {{\bf{\cW}} - {\bf{J}}} \right) = {\sigma _2}\left( {\bf{\cW}} \right),
\end{equation}
from (\ref{lemma_W_J_temp_3}), we can obtain 
\begin{align} \notag
\left| {\hat \lambda _{{\cal M},j}^{(K)} - {\lambda _{\cM}}} \right| & \le N\sum\limits_{l = 1}^{M + K} {{{\left\| {{{\left( {{\bf{\cW}} - {\bf{J}}} \right)}^{Q\left( {M + K - l + 1} \right)}}{\cbgamma ^{(l)}}} \right\|}_2}} \\ \notag
& \le N\sum\limits_{l = 1}^{M + K} {{{\left[ {{\sigma _2}\left( {\bf{\cW}} \right)} \right]}^{Q\left( {M + K - l + 1} \right)}}{{\left\| {{\cbgamma ^{(l)}}} \right\|}_2}} \\ \label{lemma_W_J_temp_4}
& \le {N^{\frac{3}{2}}}\frac{{{{\left[ {{\sigma _2}\left( {\bf{\cW}} \right)} \right]}^Q}}}{{1 - {{\left[ {{\sigma _2}\left( {\bf{\cW}} \right)} \right]}^Q}}},
\end{align} 
where (\ref{lemma_W_J_temp_4}) is from the fact that $|{\gamma_j ^{(l)}}|\le 1$ for all $j$ and $l$. This completes the proof.

\bibliographystyle{IEEEtran}
\bibliography{StochasticEncryption}

\end{document}